\documentclass[12pt]{iopart}

\usepackage{graphicx}
\usepackage[T1]{fontenc}
\usepackage{textcomp}      
\usepackage{bold-extra} 

\usepackage{cite}
\usepackage{xcolor}
\usepackage[british]{babel}
\usepackage{hyperref}
\usepackage{dirtytalk}
\usepackage{amsmath}
\usepackage{amssymb}
\usepackage[letterpaper,top=2cm,bottom=2cm,left=2cm,right=2cm,marginparwidth=1.75cm]{geometry}


\usepackage{stmaryrd}

\DeclareRobustCommand{\rchi}{{\mathpalette\irchi\relax}}
\newcommand{\irchi}[2]{\raisebox{\depth}{$#1\chi$}} 
\newcommand{\be}{\begin{equation}\begin{aligned}}
\newcommand{\ee}{\end{aligned}\end{equation}}
\usepackage{multirow}
\usepackage{booktabs}
\usepackage{textgreek}

\makeatletter
\newcommand{\showfontsize}{%
  \f@size pt%
}
\makeatother

\def\rmd{{\rm d}}

\def\rmI{{\rm I}}

\def\rmJ{{\rm J}}

\renewcommand{\vec}[1]{\boldsymbol{#1}}

\newcommand{\order}[1]{O\left(#1\right)}

\newcommand{\der}[3][\relax]{\frac{\rmd^{#1} #2}{\rmd {#3}^{#1}}}

\newcommand{\absinline}[1]{\rvert#1\rvert}

\newcommand{\pol}{\theta}
\newcommand{\tor}{{\zeta}}
\newcommand{\gyroangle}{\vartheta}

\newcommand{\grad}{{\vec{\nabla}}}
\newcommand{\gradperp}{\grad_\perp}
\ifdefined\partd
	\renewcommand{\partd}[3][]{\frac{{\partial^{#1} #2}}{{\partial #3}^{#1}}}
\else
	\newcommand{\partd}[3][]{\frac{{\partial^{#1} #2}}{{\partial #3}^{#1}}}
\fi

\newcommand{\uvec}[1]{{\vec{#1}}}

\newcommand{\vr}{{\vec{r}}}

\newcommand{\vk}{{\vec{k}}}

\ifdefined\vv
\renewcommand{\vv}{{\vec{v}}}
\else
\newcommand{\vv}{{\vec{v}}}
\fi

\newcommand{\vU}{\vec{u}}

\newcommand{\vkperp}{{\vk_\perp}}
\newcommand{\kperp}{k_\perp}
\newcommand{\kpar}{k_\parallel}

\newcommand{\vvperp}{\vv_\perp}
\newcommand{\vpar}{v_\parallel}
\newcommand{\wpar}{w_\parallel}
\newcommand{\wperp}{w_\perp}

\newcommand{\intv}[1][3]{{\int \rmd^{#1} {v} \ }}
\newcommand{\intw}[1][3]{{\int \rmd^{#1} {w} \ }}

\newcommand{\fourier}[1]{#1}

\newcommand{\s}{s}

\newcommand{\vths}[1][\s]{v_{\text{th}{#1}}}

\newcommand{\rhos}[1][\s]{\rho_{#1}}

\newcommand{\qs}[1][\s]{q_{#1}}

\newcommand{\mss}[1][\s]{m_{#1}}

\newcommand{\nss}[1][\s]{n_{#1}}

\newcommand{\Ts}[1][\s]{T_{#1}}
\newcommand{\Te}{\Ts[e]}

\newcommand{\fs}[1][\s]{f_{#1}}

\newcommand{\Fs}[1][\s]{F_{\text{M}#1}}

\newcommand{\dfs}[1][\s]{\delta \! f_{#1}}

\newcommand{\rhostar}{\rho_*}

\newcommand{\vRs}[1][\s]{{\vec{R}_{#1}}}

\newcommand{\hs}[1][\s]{h_{#1}}

\newcommand{\hskperp}[1][\s]{\fourier{h}_{{#1}\vkperp}}

\newcommand{\ess}[1][\s]{\varepsilon_{#1}}
\newcommand{\mus}[1][\s]{\mu_{#1}}

\newcommand{\vchi}{\vec{v}_{\chi}}

\newcommand{\vds}[1][\s]{\vec{v}_{\text{d}#1}}

\newcommand{\phipot}{\delta \phi}  

\newcommand{\chipot}{\chi}

\newcommand{\dApar}{\delta   A_\parallel}

\newcommand{\dAparkperp}{\delta  \fourier{A}_{\parallel\vkperp}}

\newcommand{\dBpar}{\delta   B_\parallel}

\newcommand{\dBparkperp}{\delta   \fourier{B}_{\parallel\vkperp}}

\newcommand{\vw}{{\vec{w}}}

\newcommand{\vB}{\vec{B}}
\newcommand{\ub}{\uvec{b}}

\newcommand{\vA}{\vec{A}}

\newcommand{\vdB}{\delta  \vec{B}}
\newcommand{\vdBperp}{\delta  \vec{B}_\perp}
\newcommand{\vdJ}{\delta\vec{J}}

\newcommand{\vdA}{\delta  \vec{A}}
\newcommand{\vdAperp}{\delta  \vec{A}_\perp}

\newcommand{\vrhos}[1][\s]{\vec{\rho}_{#1}}

\newcommand{\avgRs}[2][\s]{\left\langle #2 \right\rangle_{\vRs[#1]}}

\newcommand{\avgr}[1]{\left\langle #1 \right\rangle_{\vec{r}}}

\newcommand{\fsa}[2][\psi]{\left\langle #2 \right\rangle_#1}

\newcommand{\avg}[2]{\left\langle #2 \right\rangle_{#1}}

\newcommand{\LT}[1][]{L_{T_{#1}}}
\newcommand{\LTs}{\LT[\s]}
\newcommand{\LTe}{\LT[e]}
\newcommand{\LTi}{\LT[i]}

\newcommand{\Omegas}[1][\s]{\Omega_{#1}}

\usepackage{cleveref}

\crefformat{section}{section~#2#1#3}
\Crefformat{section}{Section~#2#1#3}

\crefformat{subsection}{section~#2#1#3}
\Crefformat{subsection}{Section~#2#1#3}

\crefformat{subsubsection}{section~#2#1#3}
\Crefformat{subsubsection}{Section~#2#1#3}

\crefformat{appendix}{appendix~#2#1#3}
\Crefformat{appendix}{Appendix~#2#1#3}

\crefformat{subappendix}{appendix~#2#1#3}
\Crefformat{subappendix}{Appendix~#2#1#3}

\crefformat{subsubappendix}{appendix~#2#1#3}
\Crefformat{subsubappendix}{Appendix~#2#1#3}

\crefformat{figure}{figure~#2#1#3}
\Crefformat{figure}{Figure~#2#1#3}

\crefrangeformat{figure}{figures #3#1#4--#5#2#6}
\crefmultiformat{figure}{figures~#2#1#3}{ and~#2#1#3}{, #2#1#3}{, and~#2#1#3}

\Crefrangeformat{figure}{Figures #3#1#4--#5#2#6}
\Crefmultiformat{figure}{Figures~#2#1#3}{ and~#2#1#3}{, #2#1#3}{, and~#2#1#3}

\crefformat{equation}{(#2#1#3)}
\Crefformat{equation}{Equation~(#2#1#3)}

\crefrangeformat{equation}{(#3#1#4)--(#5#2#6)}
\crefmultiformat{equation}{(#2#1#3)}{ and~(#2#1#3)}{, (#2#1#3)}{, and~(#2#1#3)}

\Crefrangeformat{equation}{Equations~(#3#1#4)--(#5#2#6)}
\Crefmultiformat{equation}{Equations~(#2#1#3)}{ and~(#2#1#3)}{, (#2#1#3)}{, and~(#2#1#3)}
\usepackage{pgfplots}
\pgfplotsset{compat=1.18}
\usepackage{import}

\makeatletter
\def\@makefnmark{%
  \hyper@linkstart{footnote}{footnote.\@thefnmark}%
  \@textsuperscript{\normalfont\@thefnmark}%
  \hyper@linkend
}
\def\@thefnmark{\arabic{footnote}}
\def\@makefntext#1{%
  \Hy@raisedlink{\hypertarget{footnote.\@thefnmark}{}}%
  \parindent 1em%
  \noindent
  \hbox to 1.8em{\hss\textsuperscript{\@thefnmark}}#1%
}
\makeatother
\newcommand{\rotation}{\Omega_\tor}

\newcommand{\rotationshear}{\omega_{\mathrm{tor}}}

\renewcommand{\s}{a}

\newcommand{\mflux}{\Pi}
\newcommand{\mfluxtot}{\mflux_{\text{tot}}}
\newcommand{\mfluxs}{\mflux_\s}
\newcommand{\mfluxschi}{\mflux_\s^{\vchi}}
\newcommand{\mfluxsphi}{\mflux_{\s}^{\phipot}}
\newcommand{\mfluxsphipar}{\mflux_{\s, \parallel}^{\phipot}}
\newcommand{\mfluxsphiperp}{\mflux_{\s, \perp}^{\phipot}}
\newcommand{\mfluxsapar}{\mflux_{\s}^{\dApar}}
\newcommand{\mfluxsaparpar}{\mflux_{\s, \parallel}^{\dApar}}
\newcommand{\mfluxsaparperp}{\mflux_{\s, \perp}^{\dApar}}
\newcommand{\mfluxsbpar}{\mflux_{\s}^{\dBpar}}
\newcommand{\mfluxsbparpar}{\mflux_{\s, \parallel}^{\dBpar}}
\newcommand{\mfluxsbparperp}{\mflux_{\s, \perp}^{\dBpar}}
\newcommand{\mfluxem}{\mflux_{\text{EM}}}
\newcommand{\mfluxemapar}{\mflux_{\s,\text{EM}}^{\dApar}}
\newcommand{\mfluxembpar}{\mflux_{\s,\text{EM}}^{\dBpar}}
\newcommand{\mfluxboot}{\mflux_{\mathrm{BS}}}
\newcommand{\mfluxgb}{\mflux_{\mathrm{gB}}}

\newcommand{\turbavg}[1]{\left< #1 \right>_{\perp}}
\newcommand{\besselarg}{a_\s}
\newcommand{\besselargint}{b_\s}

\newcommand{\dphipot}{\phipot}  
\newcommand{\dphipotkperp}{{\fourier{\dphipot}}_{\vkperp}}

\newcommand{\Hskperp}{H_{\s \vkperp}}
\renewcommand{\qs}{Z_\s e}
\newcommand{\besselgammazero}{\mathord{\text{\textGamma}}_{0\s}}
\newcommand{\besselgammaone}{\mathord{\text{\textGamma}}_{1\s}}

\newcommand{\denspseudo}{N_\s}
\newcommand{\ballooning}{\pol}
\newcommand{\bref}{B_{0}}
\newcommand{\shat}{\hat{s}}

\newcommand{\currentflux}{F_{\s}}
\newcommand{\currentbs}{J_{\mathrm{BS}}}
\newcommand{\colprefactor}{c_\nu}

\begin{document}


\title[]{Radial transport of electric current by electromagnetic microturbulence in tokamaks}

\author[]{Haomin Sun$^{1,*,\ddagger}$, Toby Adkins$^{2,\dagger,\ddagger}$, Justin Ball$^{1}$, and Yann Camenen$^{3}$}

\address{$^1$Ecole Polytechnique F\'ed\'erale de Lausanne (EPFL), Swiss Plasma Center (SPC), CH-1015 Lausanne, Switzerland}
\address{$^2$Princeton Plasma Physics Laboratory, Princeton, New Jersey, USA}
\address{$^3$Aix Marseille Univ., CNRS, PIIM, F-13397 Marseille CEDEX 20, France}
\address{$^\ddagger$These two authors contributed equally to this work}
\ead{$^{*}$haomin.sun@epfl.ch, $^\dagger$tadkins@pppl.gov}

\vspace{10pt}

\vspace{10pt}

\date{\today}

\begin{abstract}
The turbulent transport of toroidal angular momentum helps determine the rotation profiles of tokamak plasmas, and thereby their confinement and stability. The electron contribution therein has an additional consequence: even a modest electron momentum flux can correspond to a substantial turbulent flux of toroidal current, whose divergence could in principle modify the safety-factor profile. Here, using nonlinear gyrokinetic simulations, we show that electromagnetic fluctuations qualitatively alter turbulent momentum transport. In microtearing-mode-driven turbulence, the total momentum transport is inefficient relative to that of heat, yet an electron contribution associated with the turbulent Maxwell stress dominates the momentum flux. We show that this contribution exceeds an estimated scale required for turbulent current redistribution to compete with the collisional processes maintaining the bootstrap current. In the case of kinetic-ballooning-mode-driven turbulence considered, the momentum transport is found to be stronger and remains dominated by the electrostatic ion contribution; nevertheless, retaining the Maxwell stress is essential for the electron momentum flux to exceed this bootstrap-based reference scale. To enable this study, we independently implemented complete electromagnetic momentum-flux diagnostics in the gyrokinetic codes \texttt{GENE} and \texttt{CGYRO}, and verified them through linear and nonlinear cross-code benchmarks. Taken together, these results suggest that electromagnetic momentum transport may potentially be important for the coupled evolution of the rotation, current, and safety-factor profiles in high-beta tokamak plasmas.\\

\end{abstract}

\maketitle


\section{Introduction}\label{introduction}
Magnetic-confinement-fusion devices must sustain large gradients in the plasma density and temperature in order to achieve the conditions required for fusion. These same gradients, however, provide free energy for microscale instabilities, with the resultant turbulence giving rise to transport of heat, particles, and momentum across the confining magnetic field. This turbulent transport is typically the dominant source of energy loss from the plasma and therefore plays a central role in determining both the attainable equilibrium profiles and the overall performance of fusion devices \cite{Garbet2010NF,Krommes2012ARFM}. The gyrokinetic framework provides a first-principles description of such turbulence in strongly magnetised plasmas \cite{FriemanChen1982PF,Hahm1988PF,BrizardHahm2007RMP,Abel2013RPP}, and increasingly comprehensive nonlinear gyrokinetic simulations have enabled quantitative predictions of turbulent transport in experimentally relevant regimes \cite{Dimits2000PoP,Garbet2010NF,Terry2015NF,CandyWaltz2003JCP,FJenko_2001GENEcode,Goerler2011JCP,Jolliet2007CPC,Lanti2020CPC}. 

While much of this work has naturally focused on the radial transport of energy and particles, a complete description of plasma evolution also requires an understanding of turbulent transport of toroidal angular momentum. Such momentum transport is important because the resulting rotation profile can affect both the confinement and stability of tokamak plasmas. Toroidal rotation can stabilise macroscopic magnetohydrodynamic instabilities, while shear in the rotation can suppress microturbulence and thereby improve confinement \cite{RotationMHDGarofalo2002PRL,Terry2000RMP}. Since the externally applied torque available in reactor-scale plasmas is expected to be limited, the rotation profiles in future devices may be determined largely by the intrinsic sources and turbulent transport of momentum \cite{Peeters2011NF,Camenen_2011NFReview,Angioni_2012NFReview,Diamond2013NF}. Gyrokinetic theory and simulation have been used extensively to study this problem in the electrostatic limit, identifying diffusive momentum transport, momentum pinches, and residual stresses generated by a variety of symmetry-breaking mechanisms \cite{Peeters2007PRL,HahmDiamondGurcan2007PoP,Waltz2007PoP,CassonPeetersCamenen2009PoP,ParraBarnesPeeters2011PoP,ball2014,LeeBarnesParra2015PPCF,Hornsby2017momentum}. In this setting, the momentum flux is generally dominated by ions, with the electron contribution expected to be small because of the relative size of the electron mass. Considerably less is known about momentum transport in electromagnetic turbulence. At finite plasma beta --- the ratio of the thermal to magnetic pressures --- the perturbed electromagnetic fields open additional channels for momentum transport, and the connection of these terms to fluctuations in the (perturbed) parallel current means that the usual neglect of electrons can no longer be justified.

The possible importance of the electron momentum flux extends beyond its contribution to the rotation profile. A radial flux of toroidal angular momentum carried by a species is necessarily accompanied by a radial flux of the toroidal current carried by that species. Since the two differ by the species charge-to-mass ratio, an electron contribution that appears small in the angular-momentum balance can nevertheless correspond to a substantial turbulent current flux. The divergence of this flux redistributes the toroidal current density and can therefore modify the safety-factor profile. Turbulence-driven current generation and redistribution have previously been considered in electrostatic and electromagnetic settings \cite{McDevitt2013PRL,McDevitt2017PoP,Chen2021PoP,He_2018}, but their connection to the electromagnetic toroidal angular momentum flux has not been explored. A natural measure of the possible importance of this effect is provided by the bootstrap current \cite{bickerton71,Peeters2000bootstrap,Redl2021PoP}: although a turbulent current flux transports current across flux surfaces rather than directly driving a parallel current, its divergence can reinforce or oppose the processes maintaining the bootstrap current. Since the bootstrap current can constitute a substantial fraction of the total plasma current in high-performance tokamak plasmas \cite{Fujita2001PRLbootstrap,Gerhardt2011NF,Tholerus2024STEP}, turbulent current transport of a comparable magnitude could modify the current profile at leading order.

These questions become particularly important in high-beta plasmas, where electromagnetic fluctuations can qualitatively change both the unstable modes and the resulting turbulent transport, rather than merely providing a small correction to an electrostatic state \cite{adkins22,rath22}. Two important examples are microtearing modes (MTMs) and kinetic ballooning modes (KBMs). MTMs are electron-temperature-gradient-driven instabilities with tearing parity that can generate magnetic islands and stochastic magnetic-field-line transport \cite{HazeltineDobrottWang1975,DrakeLee1977PF,Gladd1980PF,Drake1980PRL}. Nonlinear gyrokinetic simulations have shown that MTM turbulence can produce substantial electron heat transport, particularly in spherical tokamaks \cite{Applegate2007PPCF,Doerk2011PRL,Guttenfelder2011PRL,Guttenfelder2012PoP056119,Roach2009PPCF}. KBMs are ion-scale, pressure-gradient-driven electromagnetic instabilities related to ideal-MHD ballooning modes \cite{Connor1978PRLBallooning,FriemanEtAl1980PF,Tang1980NF,Aleynikova2018JPP}, and can similarly produce significant transport in sufficiently high-beta plasmas \cite{Joiner2008PoP,Pueschel_2010popbeta,McKinney2021JPP,parisi23,kennedy23}. Both instabilities are of particular relevance to spherical tokamaks, which routinely access high normalised pressure, and to proposed spherical-tokamak power plants such as STEP \cite{Roach2009PPCF,Kaye2021,Tholerus2024STEP,giacomin24}. Despite extensive work on the heat transport produced by MTM- and KBM-driven turbulence, little is known about their transport of toroidal angular momentum.

In this work, we investigate the fully-electromagnetic toroidal angular momentum flux in both MTM- and KBM-driven turbulence. To enable such a study, we have independently implemented complete momentum-flux diagnostics in the local gyrokinetic codes \texttt{GENE} \cite{JenkoGENE2000,GermaschewskiGPUgene2021} and \texttt{CGYRO} \cite{Candy2016JCP}, including contributions from the perturbed electromagnetic fields that were absent from the previous implementations. Linear and nonlinear cross-code benchmarks demonstrate close agreement between the individual components of the flux. Applying these diagnostics to MTM-driven turbulence, we find that the momentum flux is dominated by the electron contribution to the turbulent Maxwell stress. The total momentum transport is nevertheless inefficient relative to the heat transport, while the electron momentum flux exceeds a reference scale based on the bootstrap current. KBM-driven turbulence exhibits a qualitatively different transport regime. Across the transition from ITG- to KBM-dominated turbulence, both the heat and total momentum fluxes increase, accompanied by a substantially larger Prandtl number, while the bulk transport remains dominated by the electrostatic ion contribution. Even so, the electron momentum flux rises sharply during this transition, almost entirely because of the Maxwell stress, and becomes large enough for the associated current transport to compete on the bootstrap-current scale.

The remainder of this paper is organised as follows. In \cref{sec_theory}, we introduce the gyrokinetic framework, present the complete expressions for the electromagnetic toroidal angular momentum flux, and relate the species momentum flux to the turbulent transport of toroidal current. In \cref{sec:benchmark_results}, we present linear and nonlinear benchmarks of the new diagnostics between \texttt{GENE} and \texttt{CGYRO}. In \cref{sec:momentum_transport_in_em_turbulence}, we apply these diagnostics to MTM- and KBM-driven turbulence. Finally, in \cref{sec_conclusions}, we summarise our results and discuss their possible implications for future spherical tokamaks. For completeness, the individual components of the complete electromagnetic toroidal angular momentum flux are derived in \cref{app:momentum_flux_expressions}.

\section{Momentum transport and the turbulent current flux}
\label{sec_theory}
After briefly summarising the equations of $\delta \!f$ gyrokinetics (\cref{sec:df_gyrokinetics}), we introduce the expressions for the toroidal angular momentum flux (\cref{sec:momentum_fluxes}) before discussing the closely related turbulent flux of electric current, together with its potential impact on the safety-factor profile (\cref{sec:current_flux_q}), a necessary background for the numerical investigation undertaken in \cref{sec:momentum_transport_in_em_turbulence}. Readers already familiar with the gyrokinetic framework may wish to skip ahead to \cref{sec:current_flux_q}, working backwards where further clarification is required.

\subsection{$\delta \! f$ gyrokinetics}
\label{sec:df_gyrokinetics}
Throughout this work, we consider axisymmetric toroidal plasmas, in which the equilibrium magnetic field can be expressed as 
\begin{align}
    \vB = I(\psi)\grad\tor + \grad\tor\times\grad\psi,
    \label{eq:vB_toroidal_decomposition}
\end{align}
where $\psi$ is the poloidal magnetic flux (divided by $2\pi$), $\tor$ is the toroidal angle, and $I$ is the poloidal current function. Fluctuations are assumed to obey the standard gyrokinetic ordering \cite{FriemanChen1982PF,SugamaNonlinearGyrokinetics1998,Abel2013RPP}:
\begin{equation}\begin{aligned}
	\frac{\omega}{\Omegas} \sim \frac{\nu_{\s b}}{\Omegas} \sim \frac{\kpar}{\kperp} \sim \frac{Z_\s e \phipot}{\Ts} \sim \frac{\dBpar}{B} \sim \frac{\absinline{\vdBperp}}{B} \sim \frac{\dfs}{\fs} \sim \frac{\rhos}{L} \equiv \rhostar \ll 1,
	\label{eq:gyrokinetic_ordering}
\end{aligned}\end{equation}
where $L$ is the typical length scale over which the plasma equilibrium varies (e.g., the minor radius of the device) and, for each species $\s$, we define its charge \(Z_\s\) in units of the proton charge $e$, mass \(\mss\), thermal speed \(\vths\), equilibrium temperature \(\Ts = \mss \vths^2/2\), gyrofrequency \(\Omegas =  Z_\s e  B / \mss c\), and gyroradius \(\rhos=\vths / |\Omegas|\). Additionally, $\omega$ is the characteristic inverse time scale of the turbulent fluctuations, $\nu_{\s b}$ is the collision frequency between species $\s$ and $b$, $\kpar$ and $\kperp$ are the characteristic wavenumbers of the fluctuations in the directions parallel and perpendicular to the equilibrium magnetic field $\vB$, respectively, \(\phipot\) is the fluctuating electrostatic potential, and \(\dBpar\) and \(\vdBperp\) are the magnetic-field fluctuations parallel and perpendicular to the mean field, respectively. 

To the lowest order in \cref{eq:gyrokinetic_ordering}, it can be shown that the plasma supports a toroidal mean flow
\begin{align}
    \vU = \rotation R^2 \grad\tor,
    \label{eq:vU_toroidal_expression}
\end{align}
where $\rotation$ is the angular velocity and $R$ is the major radial coordinate. The distribution function $\fs$ can then be expressed as the sum of some equilibrium $\Fs$ and fluctuations $\dfs$, viz., 
\begin{equation}\begin{aligned}
    \fs = \Fs + \dfs,
    \label{eq:fs_split}
\end{aligned}\end{equation}
where the latter is further split into its gyroangle dependent and independent parts as
\begin{equation}\begin{aligned}
        \dfs = -\frac{Z_\s e \phipot'(\vr, t)}{\Ts} \Fs(\vRs, \ess) + \hs(\vRs, \ess, \mus, t).
        \label{eq:dfs_hs_decomposition}
\end{aligned}\end{equation}
Here, $\vRs = \vr - \ub \times \vw / \Omegas \equiv \vr - \vrhos(\gyroangle)$ is the guiding-centre position, where $\vr$ is the particle position, $\gyroangle$ is the gyroangle, $\ub = \vB/B$, $\ess$ is the particle energy, $\mus = \mss \wperp^2/2B$ is the particle magnetic moment, $\vv$ is the particle velocity, and $\phipot'$ is the electrostatic potential in the toroidally rotating frame. We have introduced the peculiar velocity $\vw = \vv - \vU$, whose parallel component satisfies $\wpar = \sigma \sqrt{2(\ess - \mus B-{{Z_\s} e \phi_0} + {m_\s \Omega_{\zeta}^2 R^2 /2})/\mss}$, where $\sigma = \pm 1$ denotes its sign and $\phi_0$ is the part of the equilibrium electrostatic potential that vanishes under flux-surface averaging. The equilibrium distribution function
\begin{equation}\begin{aligned}
    \Fs = \frac{\denspseudo}{\left(\sqrt{\pi} \vths \right)^{3}} e^{-\ess / \Ts},
\end{aligned}\end{equation}
is a Maxwellian with `pseudo-density' $\denspseudo$ and temperature $\Ts$, the former being a flux function related to the density $\nss$ by (see, e.g., \cite{Abel2013RPP})
\begin{align}
    \denspseudo(\psi) = \nss \exp\left( \frac{Z_\s e \phi_0}{\Ts} - \frac{{m_\s} \rotation^2 R^2 }{2 \Ts} \right).
\end{align}

The gyroangle-independent piece of the fluctuating distribution function $\hs$ evolves according to the gyrokinetic equation \cite{FriemanChen1982PF,SugamaNonlinearGyrokinetics1998,Abel2013RPP}:
\begin{align}
    \left(\partd{}{t}+{\vU}(\vRs)\cdot\partd{}{\vRs}\right)\left(\hs-\frac{Z_\s e \avgRs{\chipot}}{\Ts}\Fs\right)+\left(\wpar\vec{b}+\vds+\avgRs{\vchi}\right)\cdot\partd{\hs}{\vRs} & \nonumber\\ 
     +  \left[\partd{\Fs}{\psi} + \frac{\mss \Fs}{\Ts} \left(\frac{I\wpar}{B} + \rotation R^2 \right)\der{\rotation }{\psi} \right] \avgRs{\vchi} \cdot \grad\psi & = \avgRs{C_\text{lin}[\hs]}.
    \label{eq:gk}
\end{align} 
In \cref{eq:gk}, $\avgRs{\dots}$ denotes the standard gyroaverage at fixed gyrocentre position $\vRs$, $C_\text{lin}$ is the linearised collision operator, $\vds$ are the magnetic drifts (including corrections from rotation; see, e.g., equation (111) of \cite{Abel2013RPP}), and $\vchi = (c/B) \ub \times \grad \chipot$ is the drift due to the gyrokinetic potential $\chipot = \phipot - \vv\cdot\vdA/c = \phipot' - \vw \cdot \vdA/c$, where $\vdA$ is the fluctuating magnetic vector potential and we have adopted the Coulomb gauge $\grad\cdot\delta \vA=0$.

Finally, the fields appearing in \cref{eq:gk} are related to the distribution function $\hs$ by the quasineutrality condition
\begin{equation}\begin{aligned}
    \sum_\s\left(-\frac{Z_\s^2 e^2 \nss}{\Ts}\phipot' + Z_\s e \intw \avgr{\hs}\right) = 0,
    \label{eq:quasineutrality}
\end{aligned}\end{equation}
and Amp\`ere's law
\begin{equation}\begin{aligned}
    \grad \times \vdB= \frac{4\pi}{c} \vdJ = \frac{4\pi}{c} \sum_\s Z_\s e \intw \avgr{\vw \hs},
    \label{eq:amperes_law_fullv}
\end{aligned}\end{equation}
with $\avgr{\dots}$ denoting the standard gyroaverage at constant real (particle) position.

\subsection{Complete expressions for the toroidal angular momentum flux}
\label{sec:momentum_fluxes}
The transport-timescale evolution of the toroidal angular momentum can be written as \cite{SugamaNonlinearGyrokinetics1998,Abel2013RPP}
\begin{align}
	\frac{1}{V'} \frac{\partial}{\partial t} \fsa{V'\sum_\s {m_\s n_\s} \rotation R^2} + \frac{1}{V'} \frac{\partial}{\partial \psi} \left(V' \avg{t}{\mfluxtot} \right) = S_{\Omega},
	\label{eq:momentum_transport}
\end{align}
where $V'  = \rmd V/\rmd \psi$ is the differential volume of a flux surface, $S_\Omega$ is the external source of momentum, and $\avg{t}{\mfluxtot}$ is the time average of the total turbulent toroidal angular momentum flux (henceforth `momentum flux') driven by the fluctuations \cite{SugamaNonlinearGyrokinetics1998,ParraUpDownSym2011,Abel2013RPP,ball2016a}
\begin{align}
	\mfluxtot =  \sum_\s \mfluxs = \sum_\s \mfluxschi +\mfluxem,
	\label{eq:mflux_initial}
\end{align}
where we have defined
\begin{align}
    \mfluxschi & = \fsa{\avg{\perp}{\int \rmd^3 \vw \: {m_\s} \avgr{\left[\vec{w} \cdot (R^2 \grad \tor) + \rotation R^2\right] \hs \vchi} \cdot \grad \psi}},  \label{eq:mflux_s} \\
    \mfluxem & =  - \fsa{(\grad \psi) \cdot \avg{\perp}{\frac{\vdB \vdB}{4\pi} + \frac{1}{c} \vdJ \vdA} \cdot (R^2 \grad \tor)}, \label{eq:mflux_em}
\end{align}
in which $\avg{\perp}{\dots}$ denotes an average over the turbulence spatial scales in the plane perpendicular to the equilibrium magnetic field. Given that we will only be concerned with the momentum flux generated by the turbulent fluctuations, we have neglected both the classical and neoclassical contributions to \cref{eq:mflux_initial}.

In what follows, it will be useful to separate the total momentum flux into its contributions from each of the electromagnetic fields as well as the contributions arising from particle motion parallel versus perpendicular to the magnetic field lines, viz., 
\begin{align}
    \mfluxtot^{\phipot} & = \sum_\s \mfluxsphi = \sum_\s \left( \mfluxsphipar + \mfluxsphiperp \right), \label{eq:mflux_phi}\\
    \mfluxtot^{\dApar} & = \sum_\s \left(\mfluxsapar + \mfluxemapar\right) =   \sum_\s \left( \mfluxsaparpar + \mfluxsaparperp + \mfluxemapar\right) , \label{eq:mflux_apar}\\
    \mfluxtot^{\dBpar} & = \sum_\s \left(\mfluxsbpar + \mfluxembpar\right) =   \sum_\s \left( \mfluxsbparpar + \mfluxsbparperp + \mfluxembpar\right) , \label{eq:mflux_bpar}
\end{align}
where in the final expressions we have also separated out the contributions to the fluxes arising from parallel and perpendicular motions, explicit expressions for which are given in \cref{app:momentum_flux_expressions}. Writing \cref{eq:mflux_em} as
\begin{align}
    \mfluxem = \sum_\s \left( \mfluxemapar + \mfluxembpar\right),
    \label{eq:mflux_em_components}
\end{align}
it is straightforward to show (see \cref{app:expression_for_mflux_em}) that the $\mfluxemapar$ contributions arise from the Maxwell stress associated with perpendicular fluctuations in the magnetic field, viz., the term proportional to $\vdB \vdB$. This term is usually dominated by the electron contribution and, as we shall see, can be significant at higher values of the plasma beta. The $\mfluxembpar$ contributions arise from the perpendicular transport of parallel momentum along perturbed field lines associated with the term proportional to $\vdJ \vdA$. While this contribution is often significantly smaller than that arising from the Maxwell stress, failure to retain this term can lead to a significant over-prediction of the size of $\mfluxtot^{\dBpar}$; see the paragraph following \cref{eq:mflux_bpar_em_fourier} for further discussion.

\subsection{The turbulent current flux and its effect on the safety-factor profile}
\label{sec:current_flux_q}
A radial flux of toroidal angular momentum carried by species $\s$ is necessarily accompanied by a radial flux of the toroidal current carried by that species. To leading order on a local flux surface, the two can be related approximately by\footnote{The charge-to-mass weighting in \cref{eq:radialfluxestimate} has an important consequence. While electrons usually make a negligible contribution to the total toroidal angular momentum due to their small mass, the same momentum flux can nevertheless correspond to a large current flux due to the division by mass appearing in \cref{eq:radialfluxestimate}. We will find, however, that such a compensation is not required for the electrons to contribute significantly to the momentum flux, and thus the current flux, in electromagnetic turbulence: the contributions associated with $\delta A_\parallel$ and $\delta B_\parallel$, including the Maxwell stress, are closely connected to fluctuations in the parallel current and therefore are typically dominated by electrons.}
\begin{equation}
    \currentflux \approx \frac{{Z_\s} e}{{m_\s} R_0} \mfluxs,
    \label{eq:radialfluxestimate}
\end{equation}
where $\currentflux$ denotes the radial flux of toroidal current, $\mfluxs$ is the corresponding species contribution to the total toroidal angular momentum flux [see \cref{eq:mflux_initial}], and $R_0$ is the major radius on the given flux surface. Here, and in what follows, we will use `current flux' to refer to the radial transport of the toroidal current, rather than to a radial electric current itself. 

The presence of such a current flux is important because its divergence modifies the toroidal current density profile, and thereby the {safety-factor} profile $q(\psi)$ --- a quantity of central importance for tokamak confinement and stability. The bootstrap current \cite{bickerton71} provides a familiar example of this, and in high-performance plasmas can produce a substantial fraction of the total plasma current. For example, JT-60U sustained fully non-inductive reversed-shear discharges in which approximately 80\% of the plasma current was supplied by the bootstrap current \cite{Fujita2001PRLbootstrap}, while NSTX achieved bootstrap-current fractions of approximately {50\%} \cite{Gerhardt2011NF}. Future high-beta spherical tokamaks, including STEP, are likewise designed to operate with a large bootstrap-current fraction \cite{Tholerus2024STEP}. Note that these fractions do not imply comparable fractional changes in $q$, since its value on each flux surface depends on the current enclosed within it and hence on the radial current profile. Rather, they serve to demonstrate that the bootstrap current is comparable to the current responsible for the equilibrium poloidal field and can consequently modify the internal $q$ profile at leading order.

While the turbulent current flux \cref{eq:radialfluxestimate} is not itself a bootstrap current --- the latter being a parallel-current drive, while the former transports toroidal current \textit{across} flux surfaces --- its divergence can locally reinforce or oppose the processes that maintain the bootstrap current. We therefore use the bootstrap current to construct a reference momentum-flux scale against which to measure these effects. This is not intended to define a unique threshold for dynamically important current redistribution; its advantage is that it can be estimated entirely from the equilibrium parameters and collisionality specified in a local gyrokinetic calculation. By contrast, constructing an analogous scale from the collisional friction associated with the total plasma current would require information about the global current-density profile that is not contained in a local gyrokinetic input. We define $\mfluxboot$ to be the electron momentum flux whose divergence over a radial scale of order the minor radius $a$ would balance the electron-ion collisional friction associated with the parallel bootstrap current. This balance is
\begin{align}
    \frac{\mfluxboot}{a} \sim \colprefactor R \frac{m_e\nu_{ei}}{e} \currentbs,
    \label{eq:bootstrap_momentum_balance}
\end{align}
where $\currentbs$ is the bootstrap current density, $\nu_{ei}$ is the electron-ion collision frequency, and the factor $\colprefactor$ is an order-unity coefficient appearing in the electron-ion collisional force. Combining a large-aspect-ratio estimate for the bootstrap current density [see equation (18) of \cite{He2018NF}]
\begin{align}
    \currentbs \sim 5 q \epsilon^{-1/2} \frac{c \nss[e] \Te}{B L_p},
    \label{eq:bootstrap_current_estimate}
\end{align}
in which $\epsilon = r/R$ is the inverse aspect ratio and $L_p^{-1} = |\grad \log p|$ is the pressure gradient scale length, with \cref{eq:bootstrap_momentum_balance}, and evaluating the result on a given flux surface, we obtain
\begin{align}
    \mfluxboot \sim 2.5 q \epsilon^{-1/2} \left(\frac{\nu_{ei}}{|\Omega_e|}\right) \left(\frac{a}{L_p}\right) \left(\frac{R_0}{\rho_s}\right)^2 \mfluxgb.
    \label{Eq_BS_momentumflux}
\end{align}
We have also made use of quasineutrality for a two-species electron-hydrogen plasma, $\nss[e] = \nss[i]$, and set $\colprefactor = 0.51$, the value for the Braginskii form of the electron-ion collisional force [see, e.g., equation (3.68) of \cite{ParraBraginskiiNotes}]. The gyro-Bohm reference value for the momentum flux is
\begin{align}
    \mfluxgb = n_i m_i R_0 c_s^2 \left(\frac{{\rho_s}}{R_0}\right)^2,
    \label{eq:PigBreal}
\end{align}
where $c_s = \sqrt{\Te/\mss[i]}$ and $\rho_s = c_s/|\Omegas[i]|$ are the ion-sound speed and radius, respectively. Thus, if the turbulent electron momentum flux satisfies $|\mflux_e| \gtrsim \mfluxboot$, its divergence can in principle compete with that associated with the bootstrap current, allowing it to modify the current, and therefore safety-factor, profile at the same order. We will show in \cref{sec:momentum_transport_in_em_turbulence} that once the electromagnetic contributions to the momentum flux given by \cref{eq:mflux_em} are retained, the electron momentum flux in MTM- and KBM-dominated turbulence exceeds this threshold at values of the plasma beta that are routinely accessed in high-beta devices such as spherical tokamaks.

\section{Momentum-flux benchmarks with \texttt{CGYRO} and \texttt{GENE}}
\label{sec:benchmark_results}
To facilitate such a comparison, the complete expressions \cref{eq:mflux_phi}-\cref{eq:mflux_bpar} for the momentum flux were implemented for the first time in both the gyrokinetic codes \texttt{GENE} \cite{JenkoGENE2000,GermaschewskiGPUgene2021} and \texttt{CGYRO} \cite{Candy2016JCP}. For the latter, the electromagnetic contributions \cref{eq:mflux_em_components} had not been previously implemented, whereas for the former, this was also the case for all six terms appearing in \cref{eq:mflux_apar} and \cref{eq:mflux_bpar}. This section details a cross-code linear (\cref{sec:linear_benchmark}) and nonlinear (\cref{sec:nonlinear_benchmark}) benchmark of these new momentum-flux diagnostics, thereby providing evidence of their successful implementation and increasing confidence in the physics results presented later in \cref{sec:momentum_transport_in_em_turbulence}.

Here, and in what follows, we solve the gyrokinetic equation \cref{eq:gk} in the local limit. We introduce a set of field-line-following coordinates ($x$, $y$, $z$), where $z$ is the coordinate along the equilibrium magnetic field --- here taken to be the straight-field-line poloidal angle $\ballooning$ --- while $x$ and $y$ are the (normalised) radial and binormal coordinates, respectively. Then, the equilibrium magnetic field \cref{eq:vB_toroidal_decomposition} can be written as $\vB = \bref \grad x \times \grad y$, for some reference magnetic field $\bref$. Due to assumed periodicity in the spatial domain perpendicular to $\vB$, the perturbations can be expressed in terms of the coordinates $(k_x, k_y, z)$, where $k_x$ and $k_y$ are radial and binormal wave numbers, respectively. In all simulations conducted in this paper, two kinetic species (ions and electrons) are evolved, as is required in the presence of finite electromagnetic perturbations.

\subsection{Linear benchmark}
\label{sec:linear_benchmark}
For the linear benchmark, we consider a single linear mode with a non-zero ballooning angle $\ballooning_0 = - k_{x0}/(k_y \shat)$, one of the ways in linear simulations to break the up-down symmetry of gyrokinetics and obtain a non-zero momentum flux \cite{ParraUpDownSym2011,Peeters2011NF}. We use circular flux surfaces parameterised using the Miller representation \cite{Millergeometry1998}, whose parameters can be found in \cref{tableLinear1}, alongside the (normalised) values of the equilibrium scale lengths 
\begin{align}
    L_{\Ts}^{-1} = - \frac{\rmd \log \Ts}{\rmd x}, \quad L_{\nss}^{-1} = - \frac{\rmd \log \nss}{\rmd x}.
    \label{eq:gradients}
\end{align}
The electron plasma beta was set to $\beta_e = 0.08$ to ensure that the contributions from $\dApar$ and $\dBpar$ to the momentum flux are sufficiently large. The numerical parameters used in both \texttt{CGYRO} and \texttt{GENE} can be found in \cref{tableLinear2}. Convergence checks have been performed by doubling the resolutions used in $k_x$, $z$, and the velocity-space coordinates $(n_\varepsilon, n_\xi)$ or $(n_{v_{\parallel}},n_{\mu})$. 

\begin{figure}
    \centering
    \includegraphics[width=0.69\textwidth]{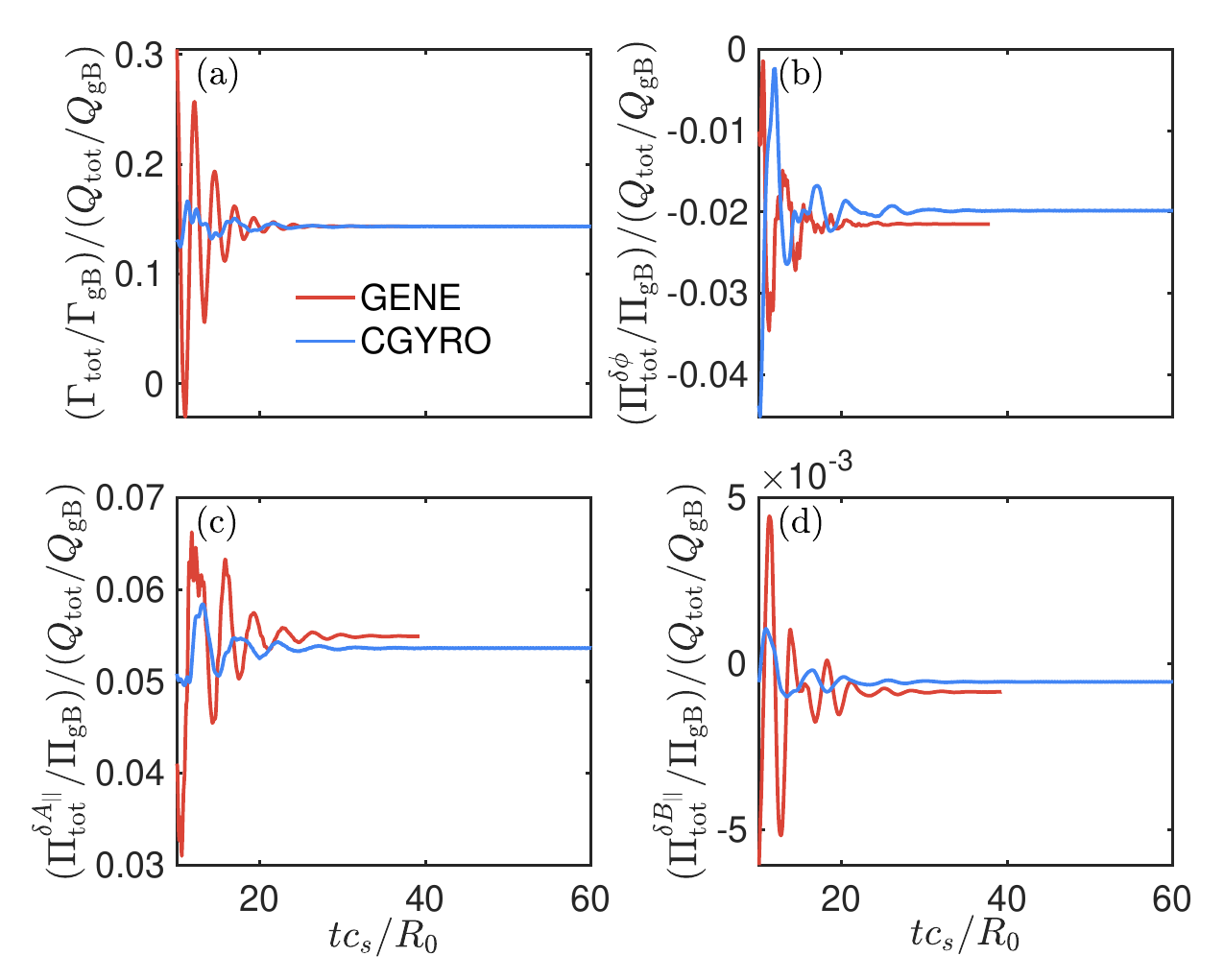}
    \caption{Results of the linear benchmark between \texttt{GENE} (red) and \texttt{CGYRO} (blue) for (a) the total particle flux and (b)-(d) different components of the toroidal angular momentum flux (normalised to the total heat flux), as defined in \cref{eq:mflux_phi}-\cref{eq:mflux_bpar}. All fluxes are normalised to their gyro-Bohm reference values of \cref{eq:PigBreal} and \cref{eq:gyrobohm_normalisations}.}
    \label{fig_L_bench1}
\end{figure}

\begin{figure}
    \centering
    \includegraphics[width=0.69\textwidth]{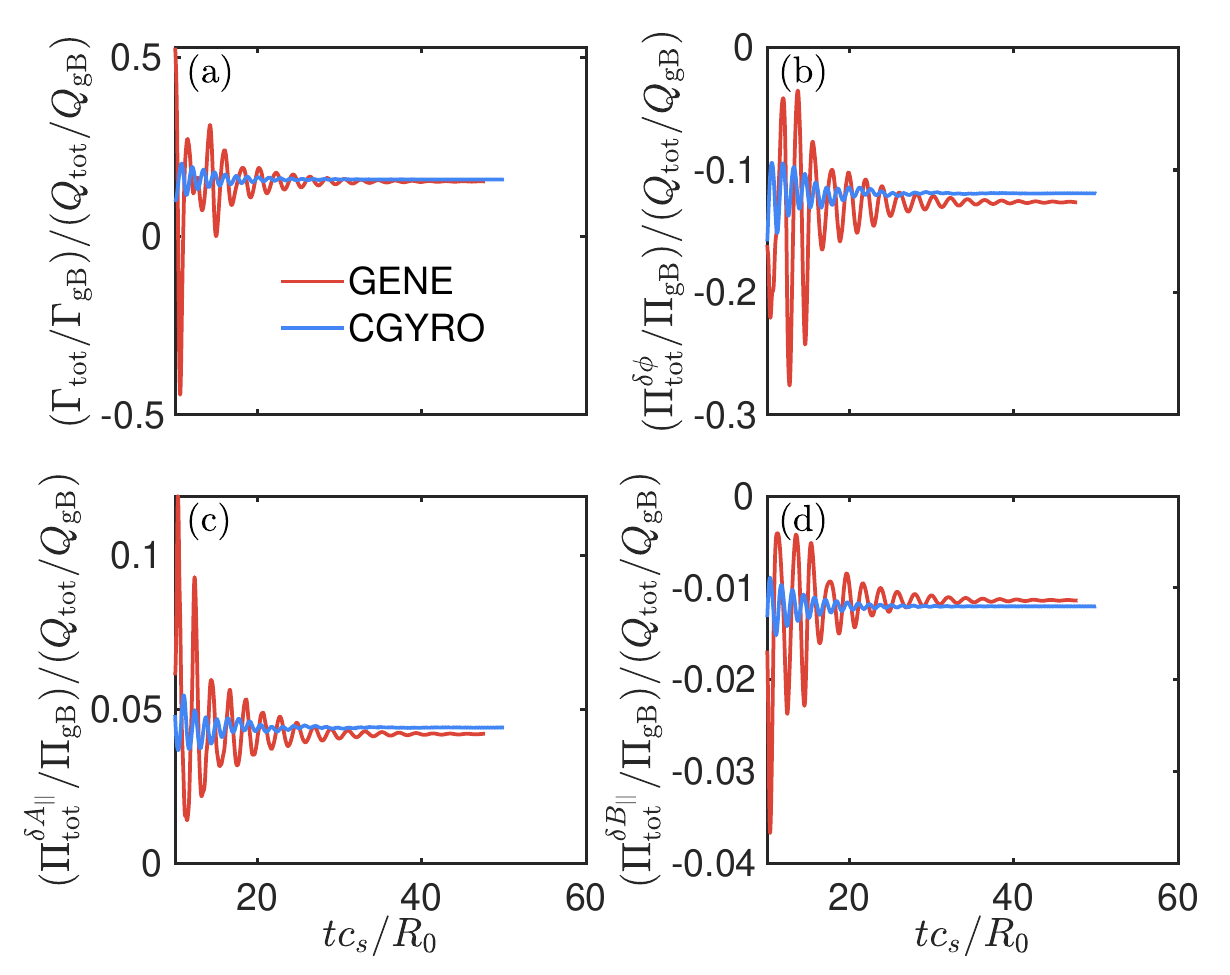}
    \caption{Same as figure \ref{fig_L_bench1}, but with $R_0/L_{Ti}$ increased from $7$ to $14$.}
    \label{fig_L_bench2}
\end{figure}

\begin{figure}
    \centering
    \includegraphics[width=0.79\textwidth]{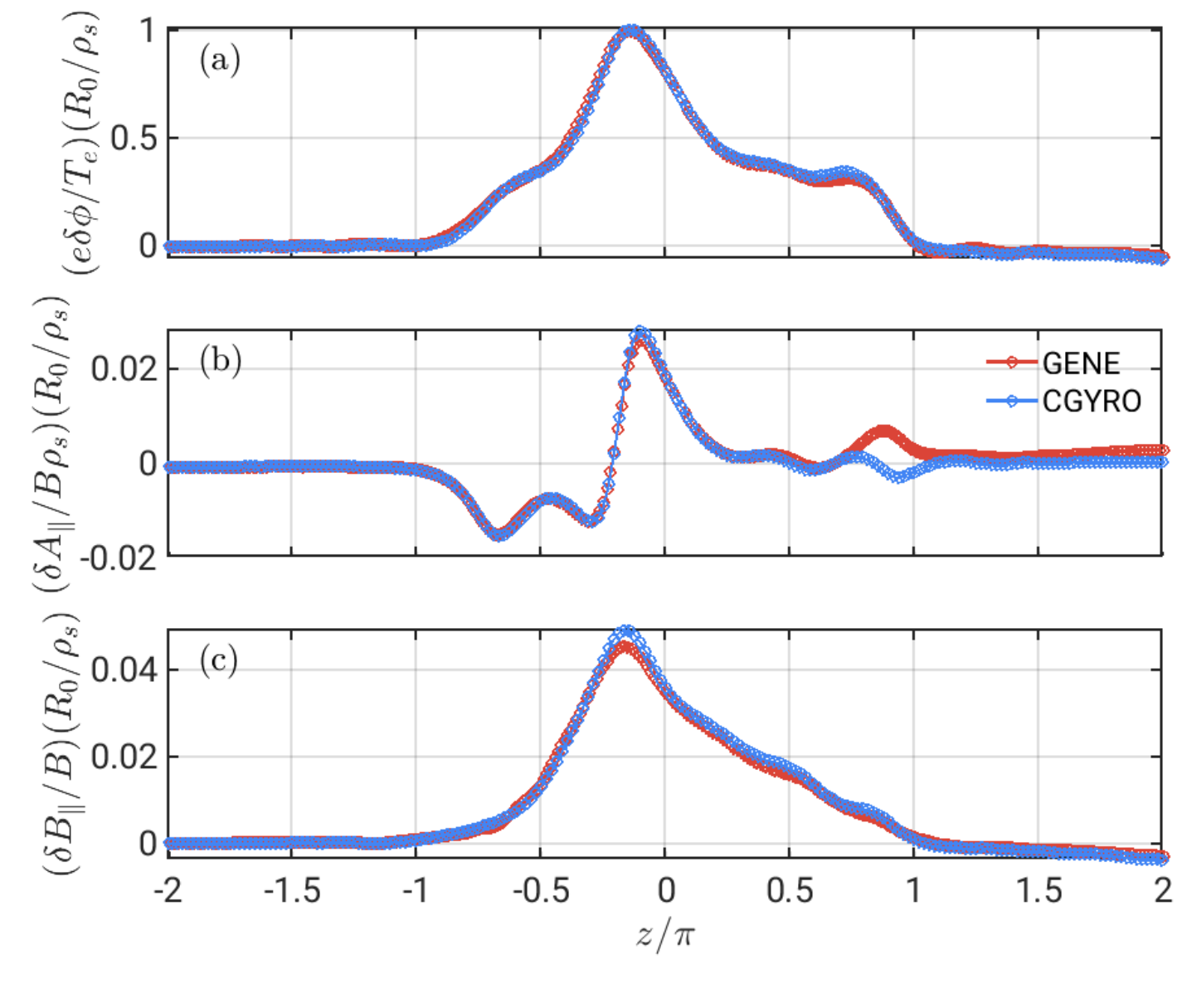}
    \caption{Comparison of the linear eigenfunctions of the perturbed (a) electrostatic potential $\delta \phi$, as well as the electromagnetic fields (b) $\delta A_{\|}$ and (c) $\delta B_{\|}$ in ballooning space (with $z = \pol$) between \texttt{GENE} (red) and \texttt{CGYRO} (blue) for the case with $R_0/L_{Ti}=14$. }
    \label{fig_L_bench_eigen}
\end{figure}

\Cref{fig_L_bench1} shows both the particle flux and the different components of the momentum flux, normalised to the total heat flux, the latter obtained by summing over all species and field components. The gyro-Bohm reference values for the particle and energy fluxes are defined as
\begin{align}
    \Gamma_{\text{gB}} = n_i c_s \left(\frac{{\rho_s}}{R_0}\right)^2, \quad  Q_{\text{gB}} = n_i T_e c_s \left(\frac{{\rho_s}}{R_0}\right)^2.
    \label{eq:gyrobohm_normalisations}
\end{align}
There is good agreement between the two codes, with less than 10\% relative error in each of the components of the momentum flux despite their small magnitude relative to the other fluxes. Nevertheless, to ensure that this remaining disagreement was due to code-specific algorithmic differences rather than issues in the implementations of the momentum-flux diagnostics themselves, we also performed an additional linear comparison at double the ion temperature gradient $R_0/\LTi$, the results of which are shown in \cref{fig_L_bench2}. The relative error becomes even smaller (less than 5\%) than in the previous case. We also compared the eigenfunctions between the two codes, which are shown in figure \ref{fig_L_bench_eigen}; all three fields show a good match throughout most of the extended ballooning space, particularly in the regions of higher amplitude.

\begin{table*}
\centering
\caption{\label{tableLinear1} Equilibrium parameters for the linear momentum-flux benchmark. Here $\shat$ is the magnetic shear, $\kappa$ and $\delta$ are the elongation and triangularity of the flux surface, and $\epsilon$ is the local inverse aspect ratio of the flux surface being simulated.}
\vspace*{0.5cm}

\begin{tabular}{ccccccccc}\hline  $\beta_e$&$R_0/L_{Ti}$&$R_0/L_{Te}$&$R_0/L_n$&$q$&$\hat{s}$& $\epsilon$&$\kappa$ & $\delta$\\ 
\hline $0.08$&$7.0$&$0$ & $1.0$&$1.4$&$0.8$&$0.36$&$1.0$ & $0.0$\\
\hline
\end{tabular}
\end{table*}

\begin{table*}
\caption{\label{tableLinear2} Numerical parameters for the linear momentum-flux benchmark of \texttt{GENE} (top) and \texttt{CGYRO} (bottom). $n_{k_x}$ and $n_{k_y}$ are the number of evolved radial and binormal wavenumbers, respectively, and $n_z$ is the number of parallel grid points. $n_{\vpar}$ and $n_\mu$ are the number of grid points in the parallel-velocity/magnetic-moment velocity representation of \texttt{GENE}, while $n_\varepsilon$ and $n_\xi$ are the number of pseudospectral grid points in the energy/pitch-angle velocity representation of \texttt{CGYRO}, with the last two columns showing the simulated ranges in these variables.}
\centering

\vspace*{0.5cm}

\begin{tabular}{l | c c c | c c c | c c c c}
\toprule
\multirow{2}{*}{\texttt{GENE}}
& $\ballooning_0$ & $k_y\rho_s$ & $z$ & $n_{k_x}$ & $n_{k_y}$ & $n_z$
& $n_{v_{||}}$ & $n_{\mu}$ & $v_{||}/\vths$ & $\mu B/{\Ts}$ \\
\cline{2-11}
& $\pi/2$ & $0.3$ & $[-\pi,\pi)$ & 48 & 1 & 96
& 128 & 18 & $[-3,3]$ & $[0,9]$ \\
\midrule
\multirow{2}{*}{\texttt{CGYRO}} 
& $\ballooning_0$ & $k_y\rho_s$ & $z$ & $n_{k_x}$ & $n_{k_y}$ & $n_z$
& $n_{\varepsilon}$ & $n_{\xi}$ & $\xi$ & $\varepsilon_\s/\Ts$ \\
\cline{2-11}
& $\pi/2$ & $0.3$ & $[-\pi,\pi)$ & 48 & 1 & 96
& 16 & 24 & [0, 1] & $[0,8]$ \\
\bottomrule
\end{tabular}

\end{table*}


\subsection{Nonlinear benchmark}
\label{sec:nonlinear_benchmark}
\begin{figure}
    \centering
    \includegraphics[width=0.55\textwidth]{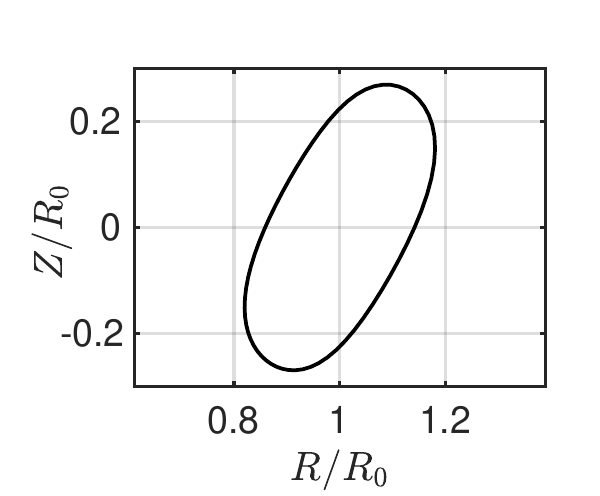}
    \caption{The up-down asymmetric flux-surface shape used for the nonlinear benchmark. $Z$ and $R$ are the vertical and radial locations of the flux surface, respectively, while $R_0$ is the value of $R$ at the centre of the flux surface.}
    \label{fig_NL_geo}
\end{figure}

In local nonlinear gyrokinetic simulations, non-zero momentum flux can be generated either by using equilibrium flow shear or by simulating turbulence in up-down asymmetric magnetic equilibria \cite{ParraUpDownSym2011,ball2018,Sun_2025_NF}. To ensure that the agreement between the two codes was not obscured by their differences in flow-shear implementation, 
we chose to use the latter in our nonlinear benchmark, such that our flux tube is located on an up-down asymmetric flux surface \cite{ball2014,Camenen_2010_TCVEXP}. To specify such a magnetic geometry, we use the extended Miller harmonic representation \cite{Arbon_2021_millermxh,Snoep2023_millermxh}, in which the shape of the flux surface is given by
\begin{align}
    R(r,\theta)&=R_0(r)+r\,\text{cos}\,\theta_R, \\
    Z(r,\theta)&=Z_0(r)+\kappa r\,\text{sin}\,\theta,
\end{align}
in which $Z$ is the vertical location of the magnetic flux surface and $R$ is the radial location of the flux surface, $\pol$ is the poloidal angle, and $\theta_R$ is
\begin{equation}
    \theta_R(\theta)=\theta+c_0+\sum^{N}_{n=1}\left[c_n\,\text{cos}(n\theta)+s_{n}\,\text{sin}(n\theta)\right].
    \label{eq:mxh_angle}
\end{equation}
For our particular case, the only non-zero contributions to \cref{eq:mxh_angle} were chosen to be $c_0=-0.589$, $c_2=-0.08$, and $s_2=0.03$, resulting in the flux-surface shape shown in \cref{fig_NL_geo}. To ensure that converged fluxes were obtained at a reasonable turbulent amplitude, a value of ${\beta_e} = 0.013$ was used to avoid the high-flux states that are often encountered in nonlinear electromagnetic gyrokinetic simulations (the so-called `non-zonal transition' \cite{pueschel13,rath22,giacomin24,zhang26fluid,zhang26gk,kennedy26}). The other equilibrium parameters can be found in \cref{tableNL1}, while the resolution parameters are given in \cref{tableNL2}. Convergence checks have been performed by doubling the box length in $x$ and $y$, which had a minimal effect on the results. 

\begin{table*}
\caption{\label{tableNL1} Equilibrium parameters for the nonlinear momentum-flux benchmark, with the definitions the same as those in \cref{tableLinear1}.}
\centering

\vspace*{0.5cm}
\begin{tabular}{ccccccccc }\hline  ${\beta_e}$&$R_0/L_{Ti}$&$R_0/L_{Te}$&$R_0/L_n$&$q$&$\hat{s}$& $\epsilon$&$\kappa$ & $\delta $ \\ 
\hline $0.013$&$9.35$&$0$ & $0$&$1.4$&$0.8$&$0.18$&$1.5$ & $0.0$ \\
\hline
\end{tabular}

\end{table*}

\begin{table*}
\caption{\label{tableNL2} Numerical parameters for the nonlinear momentum-flux benchmark of \texttt{GENE} (top) and \texttt{CGYRO} (bottom). $L_x$ and $L_y$ are the sizes of the simulation domain in the radial and binormal directions, respectively, while all other definitions are the same as those in \cref{tableLinear2}.}
\centering

\vspace*{0.5cm}

\begin{tabular}{l | c c c | c c c | c c c c}
\toprule
\multirow{2}{*}{\texttt{GENE}}
& $L_x/\rho_s$ & $L_y/\rho_s$ & $z$
& $n_{k_x}$ & $n_{k_y}$ & $n_z$
& $n_{v_{||}}$ & $n_{\mu}$ 
& $v_{||}/\vths$ & $\mu B/\Ts$ \\
\cline{2-11}
& 206 & 157 & $[-\pi,\pi)$
& 192 & 64 & 64
& 64 & 8 & $[-3,3]$ & $[0,9]$ \\
\midrule
\multirow{2}{*}{\texttt{CGYRO}}
& $L_x/\rho_s$ & $L_y/\rho_s$ & $z$
& $n_{k_x}$ & $n_{k_y}$ & $n_z$
& $n_{\varepsilon}$ & $n_{\xi}$ & $\xi$ & ${\varepsilon_\s}/\Ts$  \\
\cline{2-11}
& 157 & 157 & $[-\pi,\pi)$
& 192 & 64 & 32
& 8 & 16 & [0, 1] & $[0,8]$  \\
\bottomrule
\end{tabular}
\end{table*}

\begin{figure}
    \centering
    \includegraphics[width=0.94\textwidth]{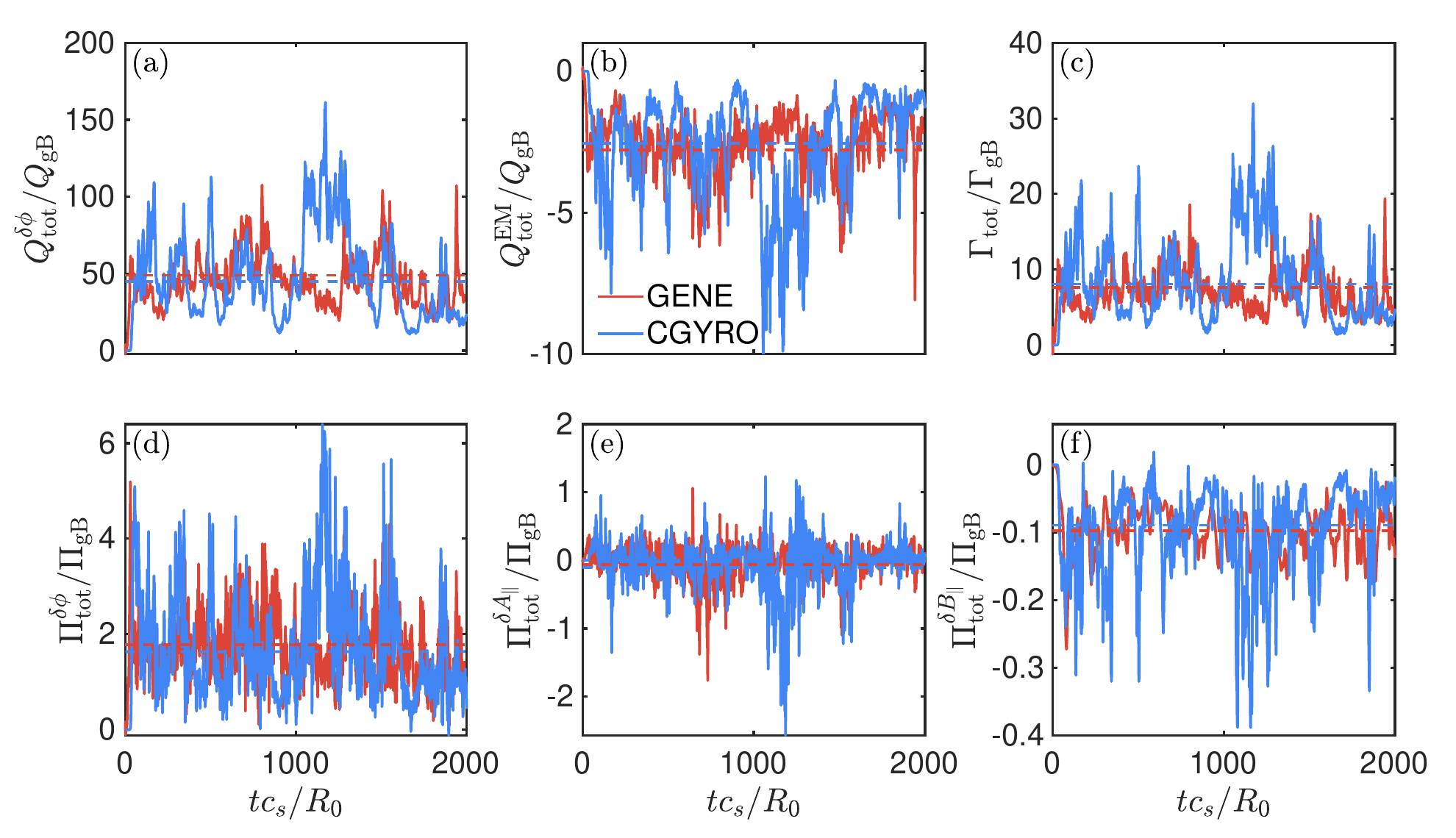}
    \caption{Results of the nonlinear benchmark between \texttt{GENE} (red) and \texttt{CGYRO} (blue) for the (a) electrostatic heat flux, (b) electromagnetic heat flux (including contributions from both $\dApar$ and $\dBpar$), (c) particle flux, (d) electrostatic component of the momentum flux \cref{eq:mflux_phi}, (e) $\dApar$ component of the momentum flux \cref{eq:mflux_apar}, and (f) $\dBpar$ component of the momentum flux \cref{eq:mflux_bpar}. All fluxes are normalised to their gyro-Bohm reference values, and the horizontal dashed lines indicate the time-averaged values over the interval $tc_s/R_0 \geqslant 200$.}
    \label{fig_NL_Benchmark}
\end{figure}

\Cref{fig_NL_Benchmark} shows the result of this nonlinear benchmark for the different fluxes, viz., those of particle, heat, and momentum. The time-averaged steady-state levels of all the fluxes show excellent agreement between the two codes, with all differences being within 5\% relative error. This, combined with the linear benchmark, suggests that the full electromagnetic toroidal angular momentum flux diagnostics have been implemented correctly in both \texttt{GENE} and \texttt{CGYRO} and can be reliably used in gyrokinetic simulations, as in the following section. To reduce computational cost, the remaining simulations in this paper were performed with only \texttt{GENE}, rather than both codes simultaneously.

\section{Toroidal angular momentum transport in MTM- and KBM-driven turbulence}
\label{sec:momentum_transport_in_em_turbulence}
Motivated by the arguments of \cref{sec:current_flux_q}, we now consider the momentum transport in MTM- and KBM-driven turbulence. All of the following simulations were conducted using a Miller representation \cite{Millergeometry1998} of a local flux surface. Given that such a representation is up-down symmetric by definition, non-zero momentum flux is created using either non-zero toroidal rotation shear $\rotationshear R_0/c_s$\footnote{Note that the rotation shear $\rotationshear$ is composed of both the gradient of the perpendicular flow \mbox{$\omega_\perp = R^2 (\grad \tor \cdot \grad y) \rmd \rotation/\rmd x$} and the gradient of the parallel flow $\omega_\parallel$. Throughout this paper, the relative amplitude of these two components is chosen self-consistently such that the total flow on each flux surface is purely toroidal. We have verified that the so-called parallel-velocity-gradient (PVG) instability \cite{CattoPVG1973,NewtonFlowShearUnderstanding2010,SchekochihinFlowShear2012} is not unstable for the values of $\rotationshear$ considered here.} or non-zero toroidal Mach number $\rotation R_0/c_s$. We will find that in sufficiently electromagnetic turbulence, the electron contribution to the momentum flux arising from the Maxwell stress is significant and comparable to the estimate \cref{Eq_BS_momentumflux} from the bootstrap current, a result that appears to be independent of the source of the momentum flux (whether rotation shear or toroidal Mach number).

\subsection{MTM-driven turbulence}
\label{MTM_NLsim}
\begin{figure}
    \centering
    \includegraphics[width=0.69\textwidth]{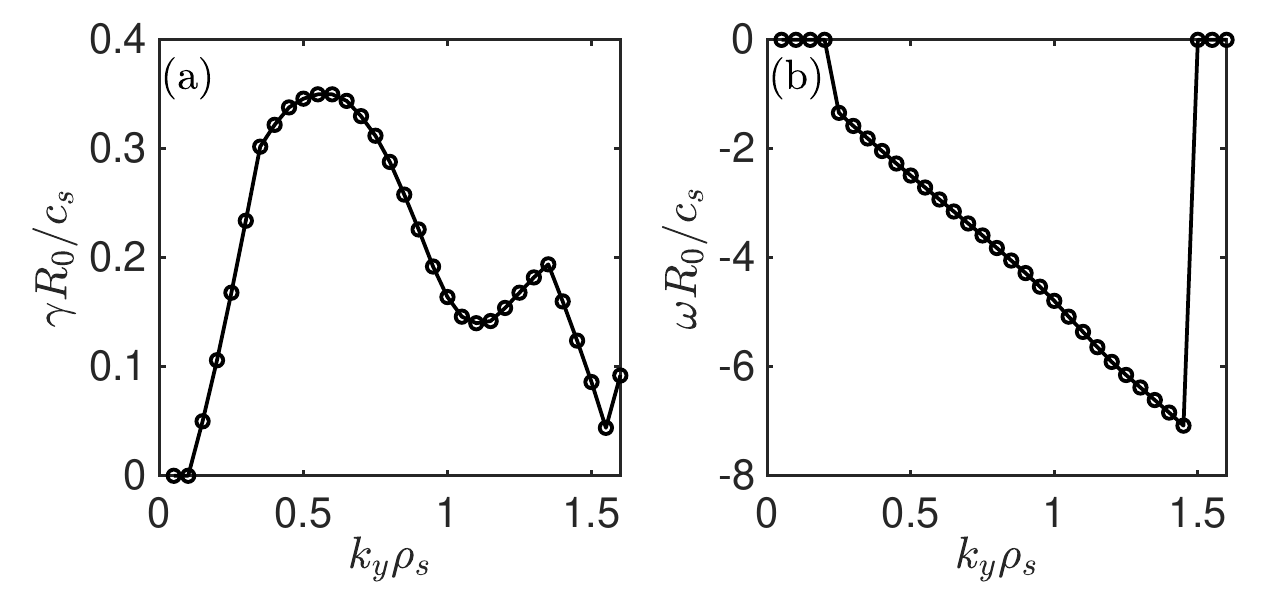}
    \caption{Linear results for the (a) growth rate and (b) real frequency as a function of $k_y\rho_s$ for ${\beta_e} = 0.03$ and $q=2$. The sign convention used for the frequency is such that positive frequencies correspond to propagation in the ion diamagnetic direction.}
    \label{fig_L_MTM_GR}
\end{figure}

\begin{figure}
    \centering
    \includegraphics[width=0.69\textwidth]{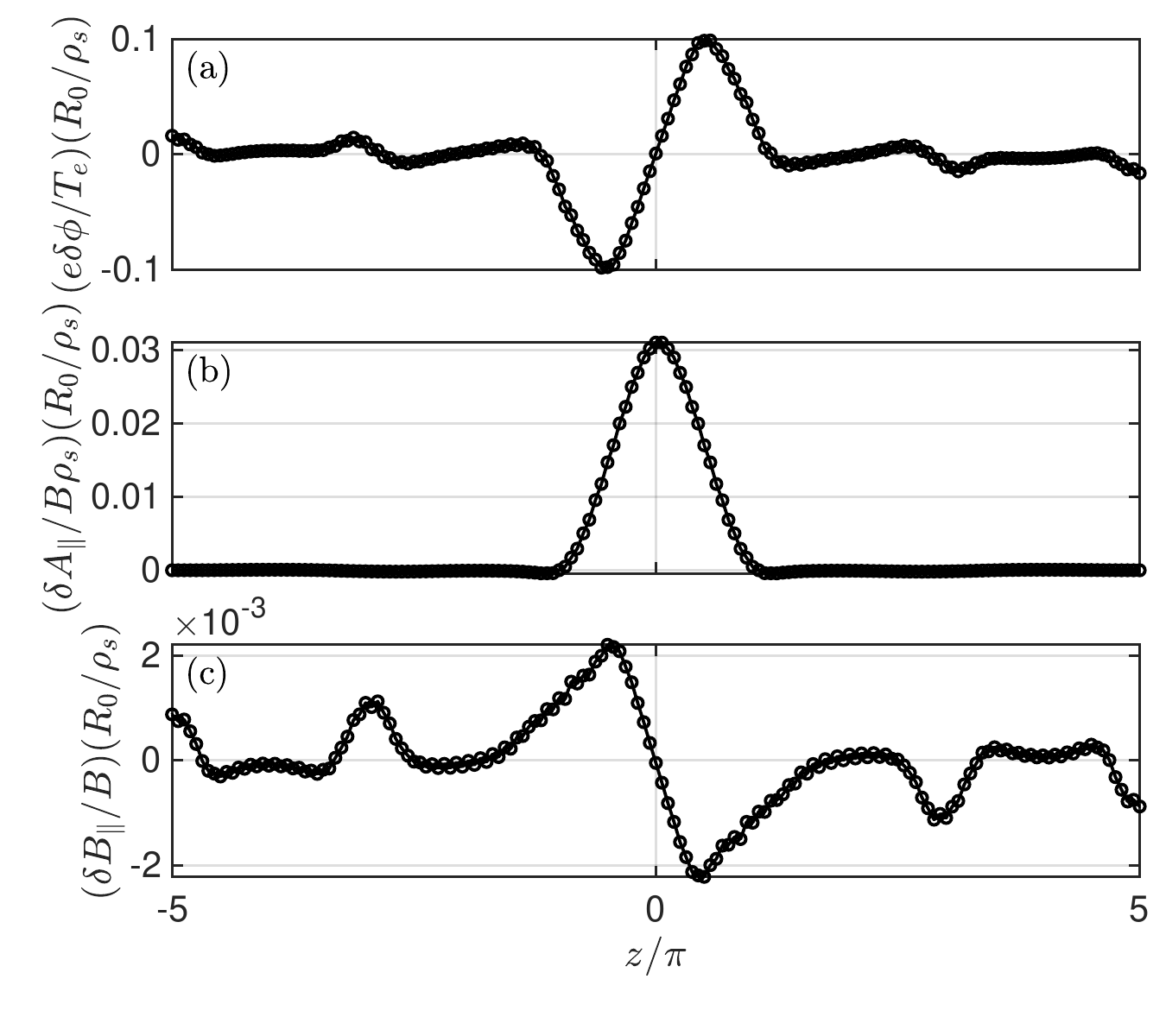}
    \caption{Eigenfunctions of the (a) electrostatic potential, as well as the electromagnetic fields (b) $\delta A_{\|}$ and (c) $\delta B_{\|}$ as a function of $z = \ballooning$ for the $k_y\rho_s=0.55$ point from figure \ref{fig_L_MTM_GR}.}
    \label{fig_L_MTM_field}
\end{figure}

\begin{figure}
    \centering
    \includegraphics[width=0.89\textwidth]{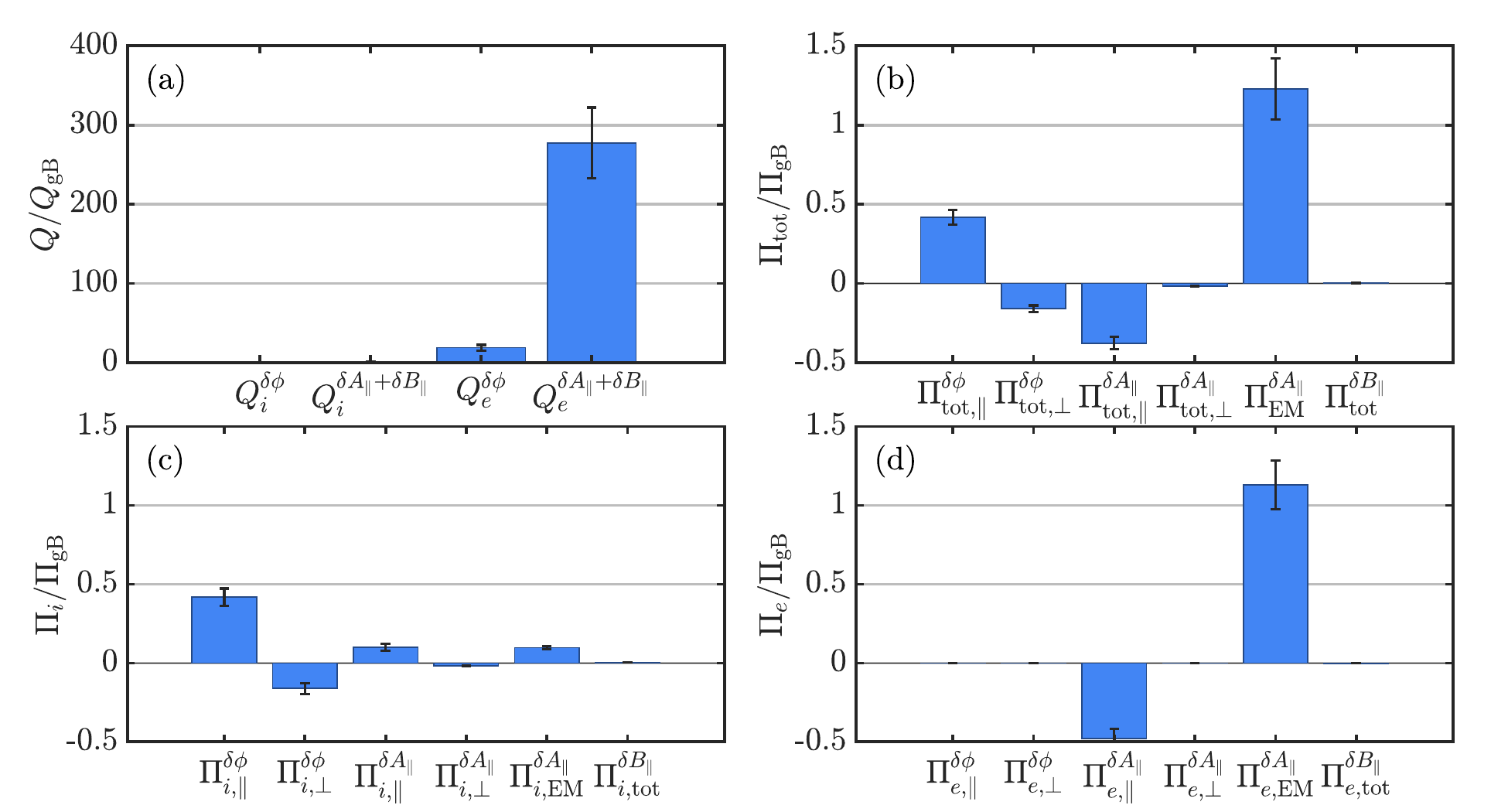}
    \caption{Contributions from the different components of the time-averaged fluxes of (a) heat, (b) total toroidal angular momentum, (c) ion toroidal angular momentum, and (d) electron toroidal angular momentum in the saturated state of the MTM-dominated case with ${\beta_e}=0.03$, $q=2$, and $\rotationshear R_0/c_s=0.12$. In (a), $Q^{\dApar+\dBpar}_i$ and $Q^{\dApar+\dBpar}_e$ are the total electromagnetic components of the ion and electron heat fluxes, respectively. The error bars denote the standard deviation in the fluxes during the saturated state.}
    \label{fig_NL_MTM_Bar}
\end{figure}

We consider dynamics on an idealised circular flux surface with a tight-aspect ratio, whose parameters can be found in \cref{tableNLMTM1}. In order to resolve the fine ballooning-space structure associated with MTMs, a large radial and velocity-space resolution is used for both the linear and nonlinear cases (see \cref{tableNLMTM2}). A linear simulation at ${\beta_e} = 0.03$ and $q=2$ confirms that the dominant instability is indeed the microtearing mode. The frequencies of the unstable modes are in the electron diamagnetic direction [see \cref{fig_L_MTM_GR}(b)], while the $\dApar$ eigenfunction [see \cref{fig_L_MTM_field}(b)] displays an even (tearing) parity typical of microtearing modes \cite{drake80,guttenfelder12,dickinson13,moraldi13,giacomin23,giacomin24,patel25}. The nature of the turbulence as being MTM-dominated is also clear from the transport characteristics in a nonlinear simulation, for which the rotation-shear was set to $\rotationshear R_0 / c_s =0.12$. In \cref{fig_NL_MTM_Bar}, we plot the time-averaged contributions to the overall heat and momentum fluxes in the saturated state. It is clear from panel (a) that the heat flux is almost entirely carried by electrons, the electromagnetic components of which are dominant, consistent with MTM-driven turbulence \cite{giacomin23,patel25}.


\begin{table*}
\centering

\caption{\label{tableNLMTM1} Equilibrium parameters for the nonlinear MTM simulations. All definitions are the same as those in \cref{tableLinear1}.}

\vspace*{0.5cm}

\begin{tabular}{cccccccccc}\hline  ${\beta_e}$&$R_0/L_{Ti}$&$R_0/L_{Te}$&$R_0/L_n$&$q$&$\hat{s}$& $\epsilon$&$\kappa$ & $\delta$\\ 
\hline $0.005-0.035$&$2$&$5$ & $1.2$&$1-4$&$2.5$&$0.36$&$1.0$ & $0.0$\\
\hline
\end{tabular}

\end{table*}

\begin{table*}
\caption{\label{tableNLMTM2} Numerical parameters for the linear (top) and nonlinear (bottom) MTM simulations, with the definitions being the same as those in \cref{tableLinear2} and \cref{tableNL2}. For the nonlinear simulations, convergence checks have been performed by doubling $N_x$, $N_y$, and the box size in $x$ and $y$, which had a negligible impact on the time-averaged fluxes.}
\centering

\vspace*{0.5cm}

\begin{tabular}{l | c c c | c c c | c c c c}
\toprule
\multirow{2}{*}{\texttt{GENE} linear}
& $\ballooning_0$ & $k_y\rho_s$ & $z$
& $n_{k_x}$ & $n_{k_y}$ & $n_z$
& $n_{v_{||}}$ & $n_{\mu}$ & $v_{||}/\vths$ & $\mu B/{\Ts}$ \\
\cline{2-11}
& $0$ & $[0.05,1.6]$ & $[-\pi,\pi)$
& 384 & 1 & 48
& 64 & 18 & $[-3,3]$ & $[0,9]$ \\
\midrule
\multirow{2}{*}{\texttt{GENE} nonlinear}
& $L_x/\rho_s$ & $L_y/\rho_s$ & $z$
& $n_{k_x}$ & $n_{k_y}$ & $n_z$
& $n_{v_{||}}$ & $n_{\mu}$ & $v_{||}/\vths$ & $\mu B/{\Ts}$ \\
\cline{2-11}
& $233$ & $157$ & $[-\pi,\pi)$
& 384 & 64 & 48
& 64 & 18 & $[-3,3]$ & $[0,9]$ \\
\bottomrule
\end{tabular}
\end{table*}

Perhaps surprisingly, \cref{fig_NL_MTM_Bar}(b) shows that the momentum flux is, in fact, dominated by the electromagnetic component $\mfluxemapar$ arising from the Maxwell stress [the first term in \cref{eq:mflux_em}], a significant departure from the usual behaviour observed in the electrostatic limit \cite{CassonPeetersCamenen2009PoP,barnes2011}. Furthermore, this component is driven primarily by electrons [see \cref{fig_NL_MTM_Bar}(c)]; this is to be expected, however, as the electrons are responsible for driving the perpendicular magnetic-field perturbations that give rise to the Maxwell stress. The contribution from $\dBpar$ is vanishingly small, as to be expected from MTM-driven turbulence [see also the discussion following \cref{eq:mflux_bpar_em_fourier}]. Given that this will prove to be the case in all simulations considered, we have chosen to plot only the total contribution to the $\dBpar$ component of the momentum flux, viz., $\Pi_{{\s},\rm tot}^{\dBpar}=\mfluxsbparpar+\mfluxsbparperp+\mfluxembpar$ [cf. \cref{eq:mflux_bpar}].

\begin{figure}
    \centering
    \includegraphics[width=0.69\textwidth]{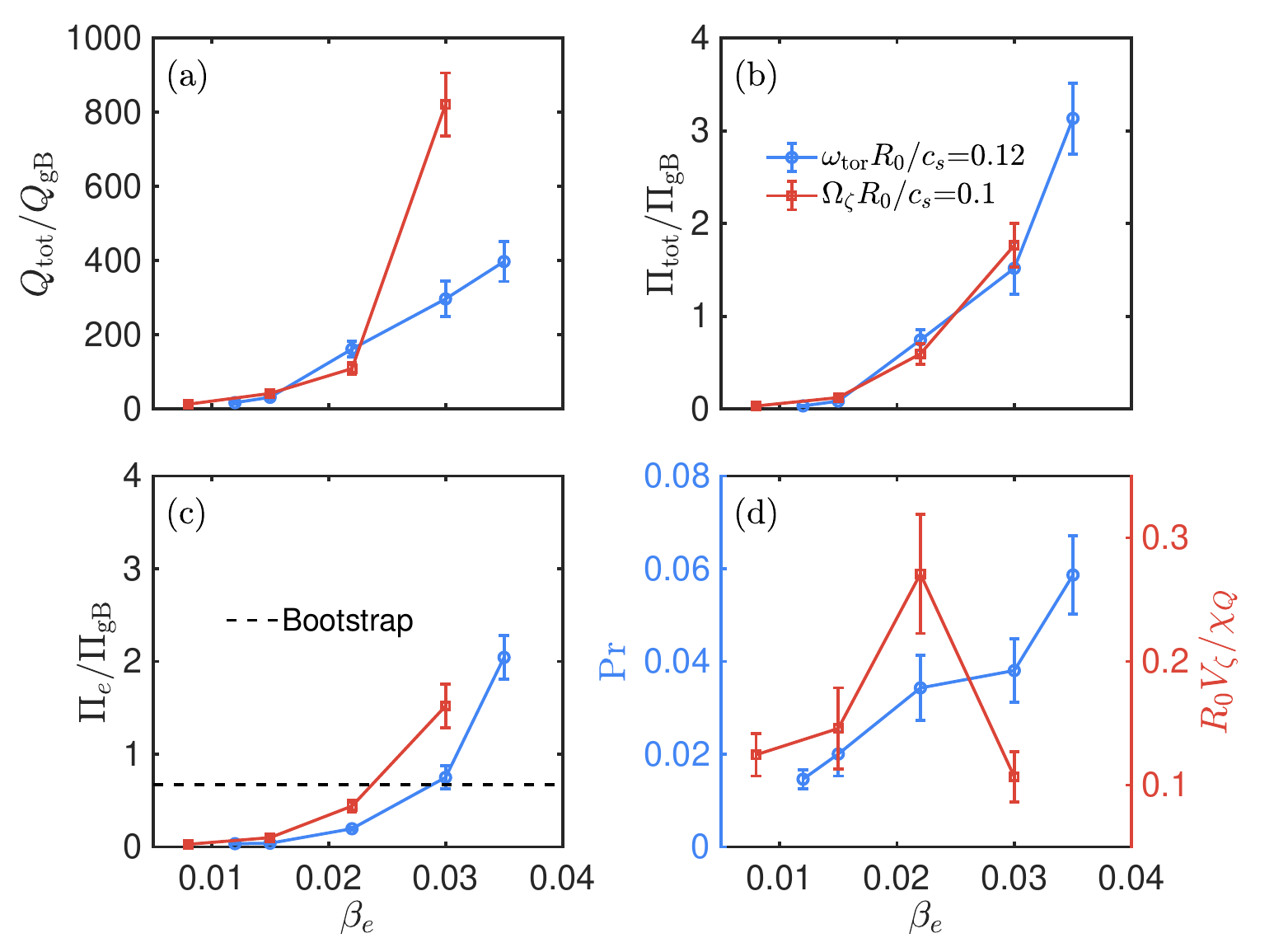}
    \caption{Time-averaged values of the (a) total heat flux, (b) total momentum flux, (c) electron momentum flux, and (d) Prandtl number $\rm Pr$ \cref{eq:Prandtlnumber_main} and pinch coefficient $R_0V_{\zeta}/\rchi_{Q}$ \cref{eq:pinch_coeff} as a function of ${\beta_e}$ for MTM-driven turbulence in the saturated state. Error bars are determined in the same way as in \cref{fig_NL_MTM_Bar}. Momentum input is supplied by either a rotation shear $\omega_{\text{tor}}R_0/c_s=0.12$ (blue circles) or non-zero toroidal Mach number $\rotation R_0/c_s=0.1$ (red squares). The horizontal dashed line in panel (c) indicates the estimate \cref{Eq_BS_momentumflux} associated with the bootstrap current.}
    \label{fig_NL_MTM_scanbeta}
\end{figure}

To assess the efficiency of the momentum transport in MTM-driven turbulence, we consider the dimensionless Prandtl number \cite{barnes2011,Highcock2011POP,ball14,Sun_2025_NF}:
\begin{align}
    \text{Pr}=\frac{Q_{\text{gB}}}{{Q}_{\text{tot}}}\frac{{\Pi}_{\text{tot}}}{\Pi_{\text{gB}}}\frac{R_0}{\LTs}\frac{\epsilon}{q}\frac{c_s}{\omega_{\text{tor}} R_0},
    \label{eq:Prandtlnumber_main}
\end{align}
which is the ratio of the momentum diffusivity to the heat diffusivity. In \cref{eq:Prandtlnumber_main}, the temperature gradient $\LTs$ is chosen to correspond to the main driving gradient of the underlying instability, with $\LTs = \LTe$ for MTM-driven turbulence. A lower Prandtl number is typically beneficial for confinement because it means that a given source of momentum (whether external or intrinsic) will drive stronger rotation shear for a given level of turbulence \cite{sun2024physicslowmomentumdiffusivity}, which can in turn reduce the other turbulent fluxes. The Prandtl number for the case reported in \cref{fig_NL_MTM_Bar} is approximately $\text{Pr}\approx0.04$ [see \cref{fig_NL_MTM_scanbeta}(d)], significantly smaller than the values observed in electrostatic ITG turbulence \cite{CassonPeetersCamenen2009PoP,barnes2011,Sun_2025_NF}. This is consistent with the fact that electron contributions dominate the transport: given their light mass, the electrons are unable to carry a large amount of momentum for a given heat flux. However, as discussed in \cref{sec:current_flux_q}, the charge-to-mass ratio appearing in the current flux \cref{eq:radialfluxestimate} suggests that this could be large for MTM-driven turbulence.

To test this hypothesis, we perform nonlinear simulations over a range of values of the plasma beta with either non-zero rotation shear $\rotationshear R_0/c_s = 0.12$ or non-zero toroidal Mach number $\rotation R_0 /c_s = 0.1$. The results of this scan are shown in \cref{fig_NL_MTM_scanbeta}. The Prandtl number remains small as ${\beta_e}$ is increased [see \cref{fig_NL_MTM_scanbeta}(d)], meaning that the ineffectiveness with which MTM-driven turbulence transports momentum is at least somewhat robust. The same is true for the so-called `pinch coefficient' \cite{Peeters2007PRLpinchterm,Tala_2011}, defined here as
\begin{align}
    \frac{R_0 V_{\zeta}}{\rchi_{Q}}=\frac{\Pi_{\mathrm{tot}}}{Q_{\mathrm{tot}}}\frac{Q_{\mathrm{gB}}}{\Pi_{\mathrm{gB}}}\frac{R_0}{\LTs}\frac{c_s}{\rotation R_0},
    \label{eq:pinch_coeff}
\end{align}
with $V_\tor$ the pinch velocity and ${\rchi_Q} = Q_\mathrm{tot} ( \nss \Ts /\LTs)^{-1}$ the heat diffusivity. \Cref{eq:pinch_coeff} quantifies the effectiveness of the toroidal rotation $\rotation$ in driving the momentum flux analogously to the Prandtl number. Crucially, however, \cref{fig_NL_MTM_scanbeta}(c) shows that the electron-driven momentum flux becomes larger than the estimate based on the bootstrap current \cref{Eq_BS_momentumflux} for values of $\beta_e \gtrsim 0.025$, indicating that the momentum transport, and thus turbulent current flux, associated with the Maxwell stress could become significant. It is worth emphasising that such a value of the plasma beta is not difficult to achieve in spherical tokamaks such as NSTX \cite{kaye19}, MAST-U \cite{MAST_Lloyd_2011} and SMART \cite{DOYLE_SMART_2021}.

\begin{figure}
    \centering
    \includegraphics[width=0.69\textwidth]{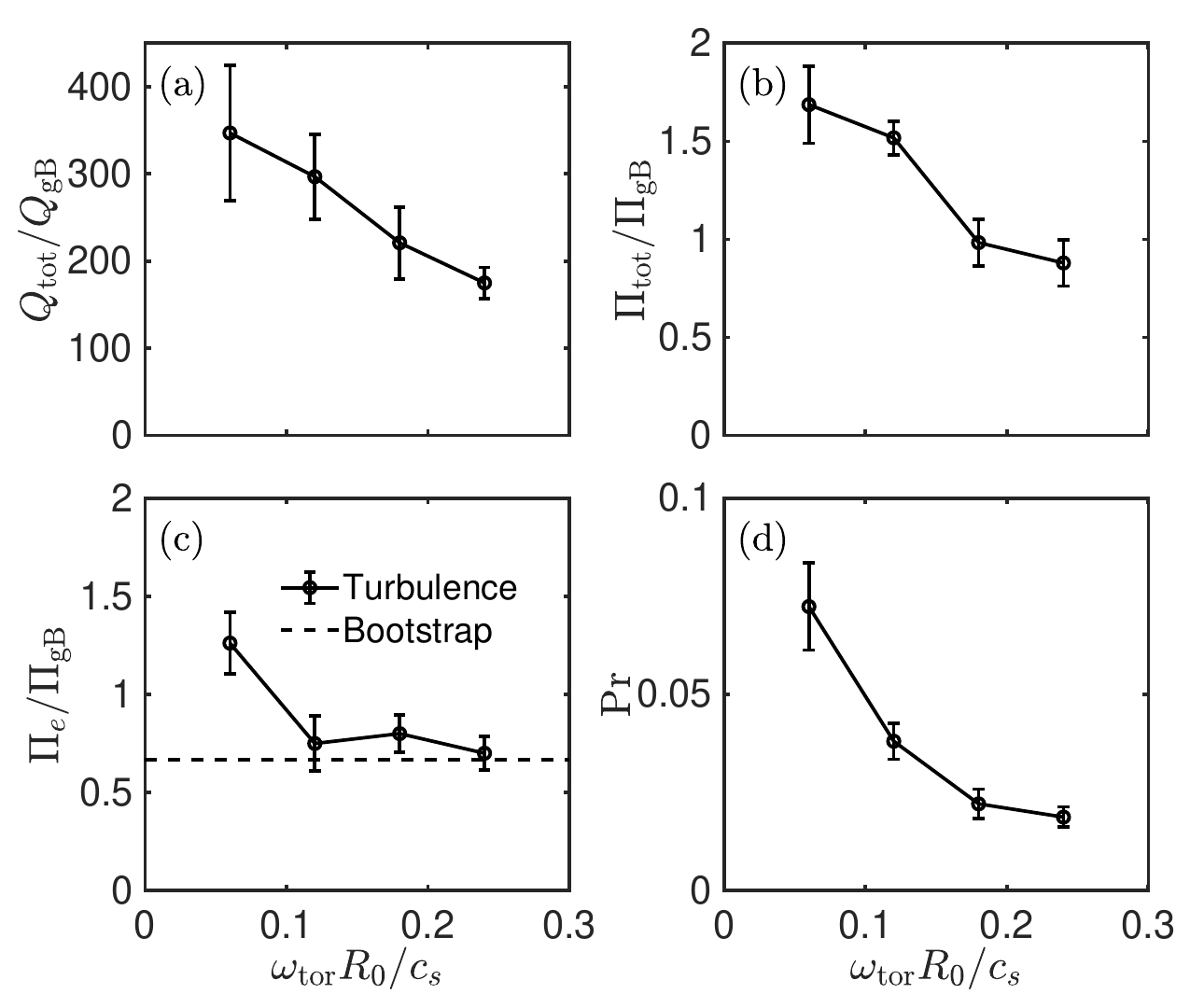}
    \caption{The same as \cref{fig_NL_MTM_scanbeta} but instead as a function of the rotation shear $\rotationshear R_0 /c_s$ and for MTM-dominated cases with ${\beta_e}=0.03$.}
    \label{fig_NL_MTM_scanExB}
\end{figure}

Finally, we perform a series of simulations with $\beta_e=0.03$ at different values of the rotation shear $\rotationshear R_0/c_s$, the results of which are shown in \cref{fig_NL_MTM_scanExB}. While the heat flux is significantly reduced with increasing rotation shear [\cref{fig_NL_MTM_scanExB}(a)], the electron momentum flux [\cref{fig_NL_MTM_scanExB}(c)] remains above the bootstrap estimate \cref{Eq_BS_momentumflux}, indicating that the significant turbulent current flux in MTM-driven turbulence is somewhat independent of the magnitude of the imposed rotation shear.

\subsection{KBM-driven turbulence}
\label{KBM_NLsim}
A natural follow-up question is whether this behaviour is unique to MTM-driven turbulence, or is rather a generic feature of all turbulence driven by predominantly electromagnetic instabilities. We thus consider KBM-driven turbulence in an equilibrium similar to that used in a previous study of finite-beta microturbulence \cite{Pueschel_2010popbeta}, for which the equilibrium and numerical parameters can be found in tables \ref{tableNLKBM1} and \ref{tableNLKBM2}, respectively.

\begin{table*}
\centering

\caption{\label{tableNLKBM1} Equilibrium parameters for the nonlinear ITG-KBM simulations. All definitions are the same as those in \cref{tableLinear1}.}

\vspace*{0.5cm}

\begin{tabular}{cccccccccc}\hline  ${\beta_e}$&$R_0/L_{Ti}$&$R_0/L_{Te}$&$R_0/L_n$&$q$&$\hat{s}$& $\epsilon$&$\kappa$ & $\delta$ \\ 
\hline $0.001-0.025$&$7$&$0$ & $1$&$1.4$&$0.8$&$0.18$&$1.0$ & $0.0$ \\
\hline
\end{tabular}

\end{table*}

\begin{table*}
\caption{\label{tableNLKBM2} Numerical parameters for the nonlinear ITG-KBM simulations. All definitions are the same as those in \cref{tableNL2}.}
\centering

\vspace*{0.5cm}

\begin{tabular}{l | c c c | c c c | c c c c}
\toprule
\multirow{2}{*}{\texttt{GENE}}
& $L_x/\rho_s$ & $L_y/\rho_s$ & $z$
& $n_{k_x}$ & $n_{k_y}$ & $n_z$
& $n_{v_{||}}$ & $n_{\mu}$ & $v_{||}/\vths$ & $\mu B/{\Ts}$ \\
\cline{2-11}
& $245$ & $157$ & $[-\pi,\pi)$
& 192 & 64 & 48
& 64 & 18 & $[-3,3]$ & $[0,9]$ \\
\bottomrule
\end{tabular}
\end{table*}

\begin{figure}
    \centering
    \includegraphics[width=0.69\textwidth]{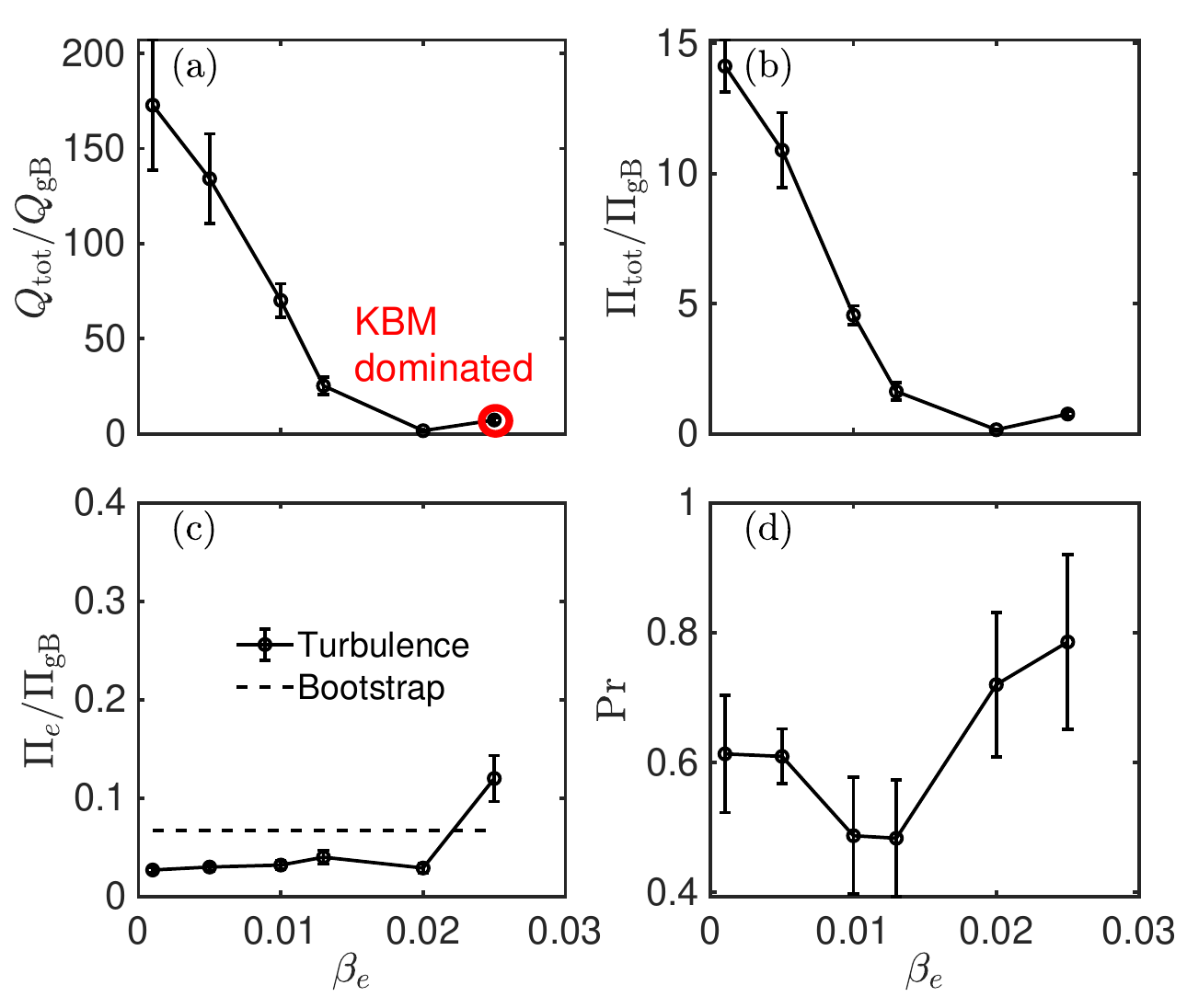}
    \caption{The same as \cref{fig_NL_MTM_scanbeta} but for the ITG-KBM cases with $\omega_{\text{tor}}R_0/c_s=0.12$. The red circle indicates that the given simulation represents KBM-driven turbulence; all other simulations are ITG-dominated.}
    \label{fig_NL_KBM_scanbeta}
\end{figure}

As for the MTM cases, we perform nonlinear simulations over a range of values of the plasma beta, the results of which are shown in figure \ref{fig_NL_KBM_scanbeta}. Only the case at the highest value of beta [${\beta_e} = 0.025$, denoted by the red circle in \cref{fig_NL_KBM_scanbeta}(a)] is KBM-dominated, with all other simulations being ITG-dominated. This is because only this case has both linear growth rates and a nonlinear heat flux that increase with beta, both of which are strong indicators of KBM-driven transport \cite{Pueschel_2010popbeta,parisi23}. Higher values of ${\beta_e}$ resulted in non-convergent nonlinear fluxes --- the so-called `non-zonal transition' \cite{pueschel13,rath22,giacomin24,zhang26fluid,zhang26gk,kennedy26} --- and we therefore focus on the difference in the transport characteristics between the ITG- and KBM-driven cases.

\begin{figure}
    \centering
    \includegraphics[width=0.89\textwidth]{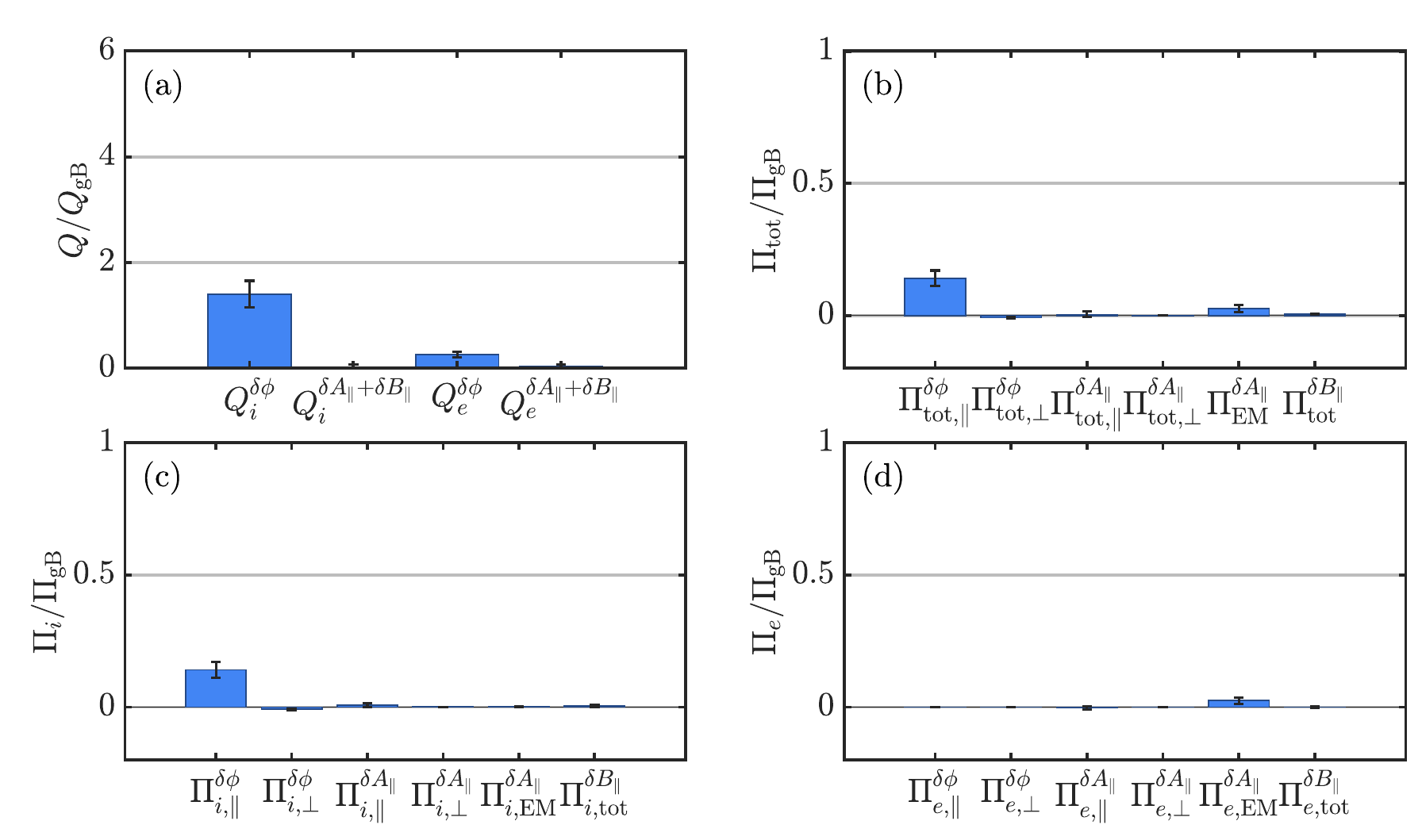}
    \caption{The same as \cref{fig_NL_MTM_Bar} but for the ITG-driven case with ${\beta_e}=0.020$.}
    \label{fig_NL_ITG_Bar}
\end{figure}

\begin{figure}
    \centering
    \includegraphics[width=0.89\textwidth]{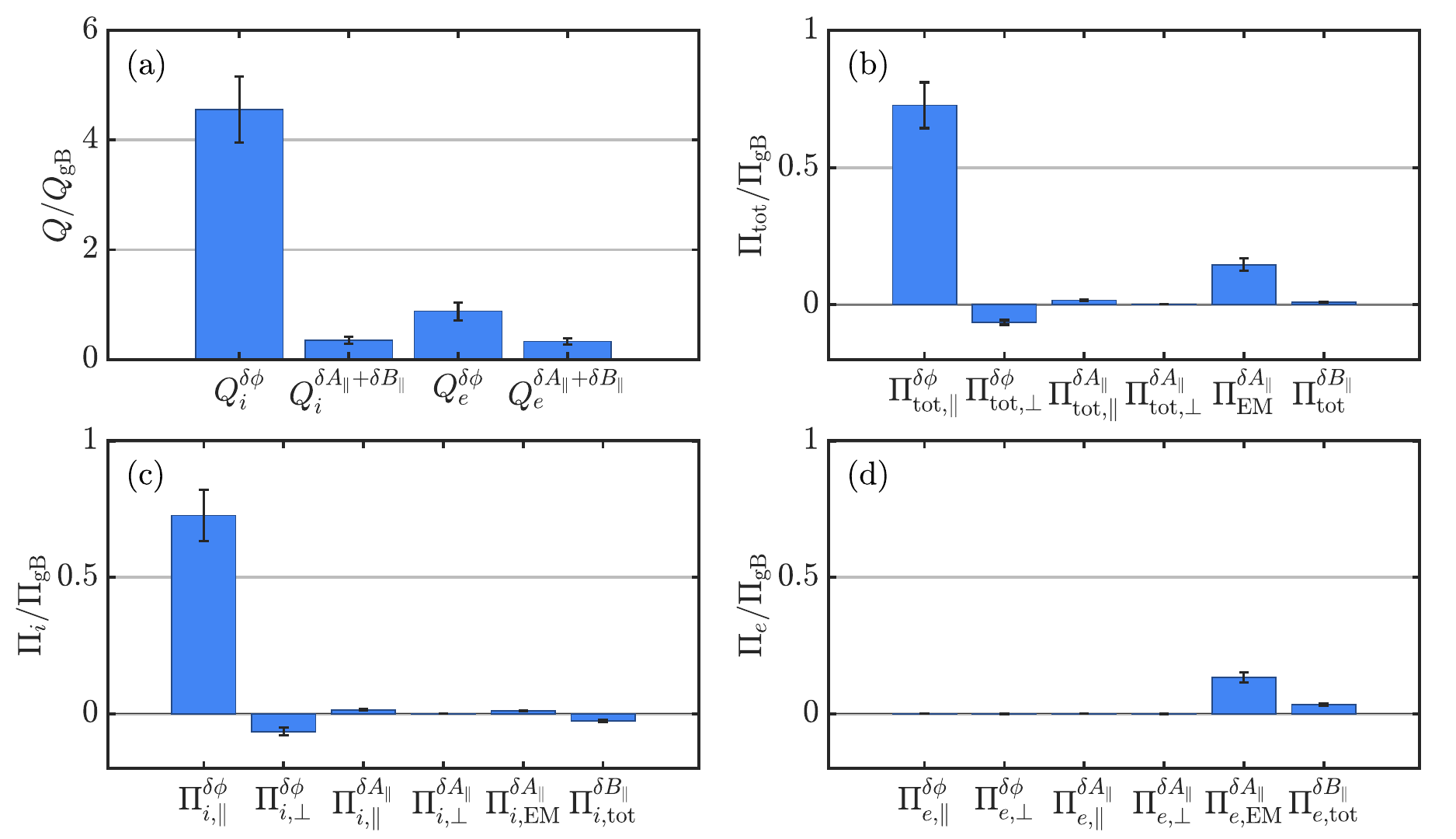}
    \caption{The same as \cref{fig_NL_MTM_Bar} but for the KBM-driven case with ${\beta_e}=0.025$.}
    \label{fig_NL_KBM_Bar}
\end{figure}

Unlike in the MTM-driven cases considered in \cref{MTM_NLsim}, the heat fluxes arising from the ITG-driven turbulence decrease as the plasma beta is increased. However, an increase in both the heat [\cref{fig_NL_KBM_scanbeta}(a)] and (total) momentum fluxes [\cref{fig_NL_KBM_scanbeta}(b)] is observed when the transition to KBM-driven turbulence occurs, which is also accompanied by an increase in the Prandtl number [\cref{fig_NL_KBM_scanbeta}(d), calculated using $\LTs = \LTi$ in \cref{eq:Prandtlnumber_main}], which has values much larger than in the MTM-driven cases considered previously. Crucially, we find that while the electron momentum flux is relatively small compared to the bootstrap current estimate \cref{Eq_BS_momentumflux} for the ITG-driven cases, it almost doubles in the transition to the KBM-driven case, becoming larger than the bootstrap estimate [\cref{fig_NL_KBM_scanbeta}(c)]. Comparing panel (d) in each of the figures \ref{fig_NL_ITG_Bar} and \ref{fig_NL_KBM_Bar}, it is clear that this increase in the electron momentum flux is due almost entirely to the component $\mfluxemapar$ associated with the Maxwell stress (both panels have the same vertical scale). Taken together, these results suggest that a large Maxwell-stress contribution to the electron momentum flux is not unique to MTM-driven turbulence. Even though the heat and total momentum transport in the KBM-driven case remain dominated by the electrostatic ion contribution, retaining the Maxwell-stress term is essential: it is this component that raises the electron momentum flux above the bootstrap-based reference scale, and hence makes the associated turbulent current flux potentially important. This conclusion is necessarily tentative, however, since it rests on a single saturated KBM-dominated simulation; a broader numerical study will be required to determine how general this behaviour is.

\section{Conclusions and discussion}
\label{sec_conclusions}
In this work, we have demonstrated that electromagnetic fluctuations can fundamentally alter the nature of turbulent momentum transport in tokamak plasmas. In particular, electrons, which are usually expected to contribute negligibly because of their small mass, can instead become dominant through their contribution to the Maxwell stress [the first term in \cref{eq:mflux_em}]. The electron momentum flux can then exceed the bootstrap-based reference scale $\mfluxboot$ \cref{Eq_BS_momentumflux}, suggesting that electromagnetic turbulence may be capable of modifying the current and safety-factor profiles at leading order \cite{McDevitt2013PRL,McDevitt2017PoP,He_2018}. To facilitate this study, we implemented the fully electromagnetic toroidal angular momentum flux diagnostics in the gyrokinetic codes \texttt{GENE} and \texttt{CGYRO}, including contributions from the perturbed electromagnetic fields that were neglected in previous implementations. The new implementations were tested through a series of linear and nonlinear cross-code benchmarks, with the time-averaged fluxes in the nonlinear case agreeing to within 5\%, providing strong evidence that both diagnostics have been implemented correctly.

Applying these diagnostics to MTM-driven turbulence, we found, perhaps surprisingly, that the momentum flux was dominated by the electron contribution to the Maxwell stress. Despite this, the total momentum flux remains small relative to the heat flux, with a Prandtl number significantly below the values typically found in electrostatic ITG-driven turbulence. A similarly small normalised pinch coefficient is obtained when the momentum flux is driven by finite toroidal rotation, even though the electron momentum flux remains large enough to exceed $\mfluxboot$. This suggests that the inefficient transport of total toroidal angular momentum is not particular to the symmetry-breaking mechanism used to generate the momentum flux. The electron momentum flux increases with the plasma beta, and exceeds the bootstrap-based reference value $\mfluxboot$ for ${\beta_e} \gtrsim 0.025$ over a range of rotation shears considered. The Maxwell-stress contribution is essential to this result: omitting it would substantially underestimate the electron momentum flux and, through \cref{eq:radialfluxestimate}, the associated turbulent current flux. MTM turbulence can thus transport toroidal angular momentum rather inefficiently while simultaneously producing a potentially significant redistribution of the toroidal current. Such a regime may allow strong rotation gradients to persist while the safety-factor profile continues to evolve under the action of turbulence.

The results obtained for KBM-driven turbulence display different transport characteristics. In the case considered here, the transition from ITG- to KBM-dominated turbulence is accompanied by an increase in both the heat and total momentum fluxes, together with a Prandtl number substantially larger than that found for MTMs. Both the heat and momentum transport in this case remain dominated by the electrostatic ion contribution. Nevertheless, the electron momentum flux increases during the transition to KBM turbulence and becomes larger than $\mfluxboot$. As in the MTM case, retaining the Maxwell-stress contribution is necessary to capture the magnitude of this electron flux. Thus, even where the total turbulent transport appears to be predominantly electrostatic and ion driven, the electromagnetic electron contribution can remain important for the transport of current. That said, given that only a single fully KBM-dominated saturated state was obtained before the onset of the non-zonal transition \cite{pueschel13,rath22,giacomin24,zhang26fluid,zhang26gk,kennedy26} often observed at higher beta, a broader numerical study will be required to determine how general this behaviour is.

These results are particularly relevant to spherical tokamaks, which access high values of the plasma beta and in which electromagnetic instabilities are already known to play an important role in turbulent transport \cite{guttenfelder12,Kaye2021,parisi23,kennedy23}. They may be even more important for future spherical-tokamak power plants such as STEP \cite{wilson20}, whose candidate operating points combine high normalised pressure with a large bootstrap-current fraction \cite{Tholerus2024STEP}, and for which electromagnetic gyrokinetic simulations predict significant transport driven by kinetic-ballooning-like turbulence \cite{giacomin24}. In such conditions, neglecting electromagnetic momentum transport could lead not only to an incorrect prediction of the rotation profile but also to a leading-order error in the evolution of the current and safety-factor profiles. Any comparison with $\mfluxboot$ made here should, however, be interpreted as establishing the potential importance of this effect: the actual modification of the current profile will depend on the sign, radial variation, and global divergence of the turbulent current flux, none of which can be determined from the local simulations presented here. Finally, we have not attempted here to provide a physical explanation as to why the Maxwell stress becomes so large in MTM- and KBM-driven turbulence, nor for the markedly different efficiencies with which these two types of turbulence transport toroidal angular momentum. The present work instead establishes the effect numerically, verifies the diagnostics required to measure it, and identifies the parameter regimes in which it may become important. Developing a physical theory of these dynamics, and determining their consequences for the global evolution of the rotation and current profiles, are the subjects of ongoing work.

\section{Acknowledgements}
The authors thank S. Brunner, B. McMillan, A. Hoffmann, F. Casson, A. Volcokas, and A. Balestri for the fruitful discussions at various stages of this project. The simulations in this work were performed on CSCS Daint and Cineca Pitagora. This work was supported by a grant from the Swiss National Supercomputing Centre (CSCS) under project IDs lp34 and lp134. This work has been carried out within the framework of the EUROfusion Consortium, partially funded by the European Union via the Euratom Research and Training Programme (Grant Agreement No. 101052200 - EUROfusion). The Swiss contribution to this work has been funded by the Swiss State Secretariat for Education, Research and Innovation (SERI). Views and opinions expressed are, however, those of the author(s) only and do not necessarily reflect those of the European Union, the European Commission, or SERI. Neither the European Union nor the European Commission nor SERI can be held responsible for them. This work was supported in part by the Swiss National Science Foundation and by the EPSRC Energy Programme (Grant Number EP/W006839/1). The work of T.A. was supported in part by the Laboratory Directed Research and Development (LDRD) Program at the Princeton Plasma Physics Laboratory for the U.S. Department of Energy under Contract No. DE-AC02-09CH11466. The United States Government retains a non-exclusive, paid-up, irrevocable, world-wide license to publish or reproduce the published form of this manuscript, or allow others to do so, for United States Government purposes.

\clearpage
\appendix

\renewcommand{\thesection}{\Alph{section}}
\renewcommand{\thesubsection}{\thesection.\arabic{subsection}}
\renewcommand{\thesubsubsection}{\thesubsection.\arabic{subsubsection}}

\crefalias{section}{appendix}
\crefalias{subsection}{subappendix}
\crefalias{subsubsection}{subsubappendix}

\section{Fourier-space expressions for the momentum flux}
\label{app:momentum_flux_expressions}
In this appendix, we derive expressions for the components of the total electromagnetic toroidal angular momentum flux $\mfluxtot$ in the Fourier-space representation (in the perpendicular spatial dimensions) that is most often used in gyrokinetic simulations. Specifically, we consider the local limit in which the gyrokinetic equation \cref{eq:gk} is solved in a domain that is asymptotically narrow compared to the variation of the plasma equilibrium. In this way, the gradients associated with the equilibrium are taken to be constant across the perpendicular domain. We can write the equilibrium magnetic field in terms of its Clebsch representation $\vB = \grad{\alpha} \times \grad{\psi}$ (in the \texttt{CGYRO} coordinate system), where $\alpha = \tor - q(\psi) \pol$, $\pol$ is the poloidal angle, and $q$ is the safety factor \cite{kruskal58,dhaeseleer91}. In these coordinates, we expand the fluctuating distribution function and electromagnetic fields in terms of their perpendicular Fourier components, viz., 
\begin{align}
	\hs = \sum_{\vkperp} e^{i\vkperp\cdot \vRs} \hskperp, \quad \chi = \sum_{\vkperp} e^{i\vkperp\cdot \vr} \chi_{\vkperp},
	\label{eq:fourier_components}
\end{align}
where $\vkperp =  k_\alpha \grad\alpha + k_\psi \grad{\psi}$ is the perpendicular wavenumber. Although the final expressions for the momentum flux that we obtain below already exist in the literature (see, e.g.,\cite{SugamaNonlinearGyrokinetics1998,ball2016a}), we find it useful to include an explicit derivation here for the sake of completeness.

The remainder of this appendix is organised as follows. We derive expressions for the species-dependent \cref{eq:mflux_s} and electromagnetic \cref{eq:mflux_em} parts of the momentum flux in \cref{app:expression_for_mflux_s} and \cref{app:expression_for_mflux_em}, respectively, which are then broken down into the contributions from $\phi$, $A_{\|}$, and $B_{\|}$ in \cref{app:contributions_by_field}, where details of the implementation in \texttt{CGYRO} are also briefly discussed. Finally, \cref{app:expressions_for_gene} details the explicit form of the expressions implemented in \texttt{GENE}. 

\subsection{Expression for the species-dependent contribution \cref{eq:mflux_s}}
\label{app:expression_for_mflux_s}
Using $\vec{w} = w_\parallel \ub + \vec{w}_\perp$ and the definition of the toroidal current function $I = R^2 \vB\cdot \grad{\tor}$, \cref{eq:mflux_s} can be written as
\begin{align}
 \mfluxschi & =  \fsa{\turbavg{\int \rmd^3 \vec{w} \: \mss \avgr{\left[\vec{w} \cdot (R^2 \grad \tor) + \rotation R^2\right] \hs \vchi} \cdot \grad \psi}} \nonumber \\
 & = \fsa{\turbavg{\int \rmd^3 \vec{w} \: \mss \avgr{\left[\left(\frac{I w_\parallel}{B} + \rotation R^2\right) + \vec{w}_\perp \cdot (R^2 \grad \tor)\right] \hs \vchi} \cdot \grad \psi}}.
 \label{eq:first_term_initial}
\end{align}
Throughout this appendix, we make extensive use of the identity
\begin{align}
	R^2 \grad \tor = \frac{I \ub }{B} - \frac{\ub \times \grad \psi}{B} \quad \Leftrightarrow \quad \ub \times \grad \psi = I \ub - BR^2 \grad \tor,
	\label{eq:magnetic_field_identity}
\end{align}
which follows directly from the definition of the axisymmetric magnetic field \cref{eq:vB_toroidal_decomposition}. Recalling the definition of $\vchi$ and \cref{eq:vB_toroidal_decomposition}, we may write
\begin{align}
	\vchi \cdot \grad \psi = \frac{c}{B}(\ub \times \grad \chi) \cdot \grad \psi = c \left(R^2 \grad \tor - \frac{I \ub }{B}\right)\cdot \grad \chi = c (R^2 \grad \tor) \cdot \grad_\perp \chi + \order{\rhostar},
\end{align}
such that \cref{eq:first_term_initial} becomes
\begin{align}
	\mfluxschi = \fsa{\turbavg{\int \rmd^3 \vec{w} \: \mss c \avgr{\left[\left(\frac{I w_\parallel}{B} + \rotation R^2\right) - \frac{(\vec{w}_\perp \times \ub) \cdot \grad \psi}{B}\right] \hs (R^2 \grad \zeta) \cdot \grad_\perp \chi}}},
	\label{eq:first_term_rewrite}
\end{align}
where we have also used $\vec{w}_\perp \cdot (R^2\grad \zeta) = -(\vec{w}_\perp \times \ub) \cdot \grad \psi/B$ to rewrite the final term in the square brackets. Expanding the distribution function and fields according to \cref{eq:fourier_components} and defining $\vrhos = \ub \times \vec{w}/\Omegas$, \cref{eq:first_term_rewrite} becomes
\begin{align}
	\mfluxschi = & \fsa{\turbavg{\int \rmd^3 \vec{w} \: \sum_{\vkperp} \mss c \left(\frac{I w_\parallel}{B} + \rotation R^2\right)i \vkperp \cdot (R^2 \grad \tor) \hskperp^*   \avgr{ \chi_{\vkperp} e^{i\vkperp \cdot \vrhos} }}} \nonumber \\
	& - \fsa{\turbavg{\int \rmd^3 \vec{w} \: \sum_{\vkperp} \frac{\mss c}{B}  i \vkperp \cdot (R^2 \grad \tor) \hskperp^* \avgr{\left[(\vec{w}_\perp \times \ub) \cdot \grad \psi\right] \chi_{\vkperp} e^{i\vkperp \cdot \vrhos} }}},
	\label{eq:first_term_final}
\end{align}
where the asterisk `*' denotes complex conjugation and we have made use of the reality condition $\hskperp^*(\vkperp)=\hskperp(-\vkperp)$. Using the identities of the gyroaverage
\begin{align}
	\avgr{e^{i \vkperp\cdot \vrhos}} &= \rmJ_0, \label{eq:bessel_identity_zeroth} \\
	\avgr{\vec{w}_\perp e^{i \vkperp\cdot \vrhos}}  & = - i \left(\ub \times \vkperp\right) \frac{w_\perp^2}{2 \Omegas} \frac{2 \rmJ_1}{\besselarg}, \label{eq:bessel_identity_first} \\
	\avgr{\vec{w}_\perp \vec{w}_\perp e^{i \vkperp\cdot \vrhos}}  & = w_\perp^2 \left[\left(\rmJ_0 - \frac{\rmJ_1}{\besselarg}\right) \frac{\left(\ub \times \vkperp\right) \left(\ub \times \vkperp\right)}{\kperp^2} + \frac{\rmJ_1}{\besselarg} \frac{\vkperp \vkperp}{\kperp^2}\right], \label{eq:bessel_identity_second}
\end{align}
in which $\rmJ_0$, $\rmJ_1$ are Bessel functions of the first kind \cite{abramowitz72} with argument $\besselarg = \kperp w_\perp/\Omegas$,
and recalling the definition of the gyrokinetic potential $\chi = \dphipot' - \vec{w} \cdot \delta \vA/c$, it is fairly straightforward to show that 
\begin{align}
	\avgr{ \chi_{\vkperp} e^{i\vkperp \cdot \vrhos} } = \rmJ_0(\besselarg) \left(\dphipotkperp' - \frac{w_\parallel \dAparkperp}{c}\right) + \frac{2 \rmJ_1(\besselarg)}{\besselarg} \frac{\Ts}{\qs} \frac{w_\perp^2}{\vths^2} \frac{\dBparkperp}{B}
	\label{eq:gyroaveraged_term_first}
\end{align}
and
\begin{align}
	& \avgr{\left[(\vec{w}_\perp \times \ub) \cdot \grad \psi\right] \chi_{\vkperp} e^{i\vkperp \cdot \vrhos} } \nonumber\\
	&= \Omegas \frac{i \vkperp \cdot \grad \psi}{\kperp^2} \left[- \besselarg \rmJ_1 \left(\dphipotkperp' - \frac{w_\parallel \dAparkperp}{c}\right) + \left(\rmJ_0 - \frac{\rmJ_1}{\besselarg}\right) \frac{2\Ts}{\qs} \frac{w_\perp^2}{\vths^2} \frac{\dBparkperp}{B} \right],
	\label{eq:gyroaveraged_term_second}
\end{align}
where we have used $\dBparkperp =i(\ub\times\vkperp)\cdot{\vdAperp}_{\vkperp}$. Substituting \cref{eq:gyroaveraged_term_first} and \cref{eq:gyroaveraged_term_second} into \cref{eq:first_term_final}, it follows that \cref{eq:mflux_s} can be written in an identical form to that found in \cite{SugamaNonlinearGyrokinetics1998} [cf. the second-last expressions in their equations (48) and (53), up to the summation over species], viz., 
\begin{align}
	\mfluxschi & = \fsa{\turbavg{ \int \rmd^3 \vec{w} \sum_{\kperp} m_\s c\left(\frac{I w_\parallel}{B} + \rotation R^2\right) i \vkperp \cdot (R^2 \grad \tor) \hskperp^* \avg{\vkperp}{\chi}}} \nonumber \\
	& \:\: + \fsa{\turbavg{ \int \rmd^3 \vec{w} \sum_{\kperp} \qs (R^2 \grad \tor) (\grad \psi) \colon \frac{\vkperp \vkperp}{\kperp^2} \hskperp^* \avg{\vkperp}{(\vec{w}_\perp \times \ub) \chi }}},
	\label{eq:mflux_incomplete}
\end{align}
where we have defined the following combinations of the electromagnetic fields (cf. equation (20) in \cite{SugamaNonlinearGyrokinetics1998})
\begin{align}
	\avg{\vkperp}{\chi} & = \rmJ_0(\besselarg) \left(\dphipotkperp' - \frac{w_\parallel \dAparkperp}{c}\right) + \frac{2 \rmJ_1(\besselarg)}{\besselarg} \frac{\Ts}{\qs} \frac{w_\perp^2}{\vths^2} \frac{\dBparkperp}{B}, \label{eq:chi_fouier} \\
	\avg{\vkperp}{(\vec{w}_\perp \times \ub) \chi } & = - \besselarg \rmJ_1(\besselarg) \left(\dphipotkperp' - \frac{w_\parallel \dAparkperp}{c}\right) + \left[\rmJ_0(\besselarg) - \frac{\rmJ_1(\besselarg)}{\besselarg}\right] \frac{2\Ts}{\qs} \frac{w_\perp^2}{\vths^2} \frac{\dBparkperp}{B}. \label{eq:chi_wperp_fourier}
\end{align}

\subsection{Expression for the electromagnetic contribution \cref{eq:mflux_em}}
\label{app:expression_for_mflux_em}
Using $\vdB = \grad \times {\vdA} = \dBpar \ub - \ub \times \grad{\dApar} + \order{\rhostar}$, the definition of the perturbed current \cref{eq:amperes_law_fullv} and repeated applications of \cref{eq:magnetic_field_identity}, \cref{eq:mflux_em} can be written as
\begin{align}
	\mfluxem & =  -\fsa{(\grad \psi) \cdot \turbavg{\frac{\vdB \vdB}{4\pi} + \frac{1}{c} \delta \vec{J} \delta \vA} \cdot (R^2 \grad \tor)} \nonumber \\
	& = -\frac{1}{4\pi} \fsa{\turbavg{ (\grad \dApar \times \ub)\cdot \grad{\psi} \left[\frac{I \dBpar}{B} - \frac{\left(\grad{\dApar} \times \ub\right) \cdot (\ub \times \grad \psi)}{B}\right]  }} \nonumber \\
	& \quad \:\: - \fsa{\turbavg{ \sum_\s \frac{\qs}{c} \int \rmd^3 \vec{w}\left[\frac{I \dApar}{B} - \frac{(\ub \times \grad \psi)\cdot \vdAperp}{B}\right] \avgr{\hs (\vec{w}_\perp \cdot \grad \psi)} }} \label{eq:em_terms_initial} \\
	& =  \frac{I}{4\pi} \fsa{\turbavg{ \sum_{\vkperp} i \vkperp \cdot (R^2 \grad \tor) \dAparkperp  (\dBparkperp)^*}} \nonumber \\
	& \quad + \frac{1}{4\pi} \fsa{\turbavg{  \sum_{\vkperp} (R^2 \grad \tor) (\grad \psi) \colon \vkperp \vkperp |\dAparkperp|^2 }} \nonumber  \\
	& \quad + I \fsa{\turbavg{ \sum_{\vkperp} i \vkperp \cdot (R^2 \grad \tor) \dAparkperp  \left(\frac{1}{B}\sum_{\s} \int \rmd^3 \vec{w} \: \frac{\mss w_\perp^2}{2} \frac{2 \rmJ_1}{\besselarg} \hskperp\right)^* }} \nonumber \\
	& \quad -  \sum_\s \fsa{\turbavg{\int \rmd^3 \vec{w} \sum_{\vkperp} i \vkperp \cdot (R^2 \grad{\tor}) \hskperp^* \frac{\mss w_\perp^2}{2} \frac{2\rmJ_1}{\besselarg}  \frac{(\ub \times \grad{\psi}) \cdot {\vdAperp}_{\vkperp}}{B} }}.\label{eq:em_terms_intermediate}
\end{align}
Note that we have used the Fourier expansions \cref{eq:fourier_components}, in addition to \cref{eq:bessel_identity_first} and \cref{eq:magnetic_field_identity}, to write the gyroaverage appearing in the second term of \cref{eq:em_terms_initial} as
\begin{align}
	\avgr{(\vec{w}_\perp \cdot \grad \psi) e^{i \vkperp \cdot \vrhos}} = - i (\ub \times \vkperp) \cdot \grad \psi \frac{w_\perp^2 }{2 \Omegas} \frac{2\rmJ_1}{\besselarg} = - \frac{c}{\qs} i\vkperp \cdot (R^2 \grad \tor) \frac{m_{\s} w_\perp^2}{2} \frac{2\rmJ_1}{\besselarg}.
\end{align}
From \cref{eq:amperes_law_fullv} and \cref{eq:bessel_identity_first}, we see that the perpendicular component of Amp\`ere's law 
\begin{align}
	\frac{B \dBparkperp}{4\pi} + \sum_{\s} \int \rmd^3 \vec{w} \: \frac{\mss w_\perp^2}{2} \frac{2 \rmJ_1(\besselarg)}{\besselarg} \hskperp = 0,
	\label{eq:perpendicular_amperes_law}
\end{align}
means that the first and third terms in \cref{eq:em_terms_intermediate} cancel exactly. Finally, using $(\ub \times \grad\psi) \cdot {\vdAperp}_{\vkperp}= - i (\vkperp \cdot \grad \psi) \dBparkperp/\kperp^2$, we find:
\begin{align}
	\mfluxem & = \frac{1}{4\pi} \fsa{\turbavg{  \sum_{\vkperp} (R^2 \grad \tor) (\grad \psi) \colon \vkperp \vkperp |\dAparkperp|^2 }} \nonumber \\
  & \quad \:\: - \sum_\s \fsa{\turbavg{ \int \rmd^3 \vec{w} \sum_{\vkperp}  \qs (R^2 \grad \tor ) (\grad \psi) \colon \frac{\vkperp \vkperp}{\kperp^2} \hskperp^* \frac{2\rmJ_1}{\besselarg} \frac{\Ts}{\qs} \frac{w_\perp^2}{\vths^2} \frac{\dBparkperp}{B} }}. \label{eq:em_terms_final}
\end{align}


\subsection{Contributions by field}
\label{app:contributions_by_field}
Combining \cref{eq:mflux_incomplete} and \cref{eq:em_terms_final}, the total momentum flux \cref{eq:mflux_initial} can be written as 
\begin{align}
    \mfluxtot & = \sum_\s\fsa{\turbavg{ \int \rmd^3 \vec{w} \sum_{\vkperp}m_{\s} c \left(\frac{I w_\parallel}{B} + \rotation R^2\right) i \vkperp \cdot (R^2 \grad \tor) \hskperp^* \avg{\vkperp}{\chi}}} \nonumber \\
	& \:\: + \sum_\s \fsa{\turbavg{ \int \rmd^3 \vec{w} \sum_{\vkperp} \qs (R^2 \grad \tor) (\grad \psi) \colon \frac{\vkperp \vkperp}{\kperp^2} \hskperp^* { \overline{\avg{\vkperp}{(\vec{w}_\perp \times \ub) \chi }}} }},
    \label{eq:mflux_tot_fourier}
\end{align}
where we have used the parallel component of Amp\`ere's law
\begin{align}
	\kperp^2 \dAparkperp = \frac{4\pi}{c} \sum_\s \qs \int \rmd^3 \vec{w} \: {\rmJ_0}(\besselarg) w_\parallel \hskperp,
	\label{eq:parallel_amperes_law}
\end{align}
to write the first term in \cref{eq:em_terms_final} in terms of a velocity-space integral over $\hskperp$, and defined [cf. \cref{eq:chi_wperp_fourier}] 
\begin{align}
    \overline{\avg{\vkperp}{(\vec{w}_\perp \times \ub) \chi }}  = - \besselarg \rmJ_1(\besselarg)\dphipotkperp' & + \left[\besselarg \rmJ_1(\besselarg) + {\rmJ_0(\besselarg)}\right]  \frac{w_\parallel \dAparkperp}{c}  \nonumber \\
	& + \left[\rmJ_0(\besselarg) - \frac{{2}\rmJ_1(\besselarg)}{\besselarg}\right] \frac{2\Ts}{\qs} \frac{w_\perp^2}{\vths^2} \frac{\dBparkperp}{B}.
	\label{eq:chi_wperp_fourer_modified}
\end{align}
while the definition of $\avg{\vkperp}{\chi}$ remains \cref{eq:chi_fouier}.
Equation \cref{eq:mflux_tot_fourier} is the form of the momentum flux that has been implemented in \texttt{CGYRO} in commit \texttt{2567ac987fcfe9a0f39cc54ad6107887eb9506a0} from \url{https://github.com/gafusion/gacode} (see also \cite{belli18}), with \cref{eq:chi_fouier} and \cref{eq:chi_wperp_fourer_modified} being stored in normalised units as \texttt{jvec\_c} and \texttt{jxvec\_c}, respectively (see \texttt{cgyro/src/cgyro\_init\_arrays.F90}). Further information about the implementation can be found at \url{https://github.com/gafusion/gacode/pull/474}.

For our purposes here, it will be useful to separate the total momentum flux into its contributions from each of the electromagnetic fields according to \cref{eq:mflux_phi}-\cref{eq:mflux_bpar}. Introducing the `contravariant' wavenumber $k^\psi = \vkperp \cdot \grad \psi$, it follows straightforwardly from \cref{eq:mflux_tot_fourier} that these components can be written as follows: for the electrostatic potential $\phipot$
\begin{align}
    \mfluxsphipar & = +\fsa{\turbavg{ \int \rmd^3 \vec{w} \sum_{\vkperp}m_{\s} c \left(\frac{I w_\parallel}{B} + \rotation R^2\right) i k_\alpha \hskperp^* \rmJ_0(\besselarg) \dphipotkperp' }}, \label{eq:mflux_phi_par_fourier} \\
    \mfluxsphiperp & = -\fsa{\turbavg{ \int \rmd^3 \vec{w} \sum_{\vkperp} \qs \frac{k_\alpha k^\psi}{\kperp^2} \hskperp^* \besselarg \rmJ_1(\besselarg)\dphipotkperp' }} ; \label{eq:mflux_phi_perp_fourier}
\end{align}
the parallel magnetic-vector potential $\dApar$
\begin{align}
    \mfluxsaparpar & = -\fsa{\turbavg{ \int \rmd^3 \vec{w} \sum_{\vkperp} m_{\s} c \left(\frac{I w_\parallel}{B} + \rotation R^2\right) i k_\alpha \hskperp^*  \rmJ_0(\besselarg) \frac{w_\parallel \dAparkperp}{c} }}, \label{eq:mflux_apar_par_fourier}\\
    \mfluxsaparperp & = +\fsa{\turbavg{ \int \rmd^3 \vec{w} \sum_{\vkperp} \qs \frac{k_\alpha k^\psi}{\kperp^2} \hskperp^*  \besselarg \rmJ_1(\besselarg) \frac{w_\parallel \dAparkperp}{c} }}, \label{eq:mflux_apar_perp_fourier}\\
    \mfluxemapar & = +\fsa{\turbavg{ \int \rmd^3 \vec{w} \sum_{\vkperp} \qs \frac{k_\alpha k^\psi}{\kperp^2} \hskperp^*  \rmJ_0(\besselarg) \frac{w_\parallel \dAparkperp}{c} }}; \label{eq:mflux_apar_em_fourier}
\end{align}
and the parallel magnetic-field perturbations $\dBpar$
\begin{align}
    \mfluxsbparpar & = +\fsa{\turbavg{ \int \rmd^3 \vec{w} \sum_{\vkperp}m_{\s} c \left(\frac{I w_\parallel}{B} + \rotation R^2\right) i k_\alpha \hskperp^*  \frac{2 \rmJ_1(\besselarg)}{\besselarg} \frac{\Ts}{\qs} \frac{w_\perp^2}{\vths^2} \frac{\dBparkperp}{B} }}, \label{eq:mflux_bpar_par_fourier}\\
    \mfluxsbparperp & = +\fsa{\turbavg{ \int \rmd^3 \vec{w} \sum_{\vkperp} \Ts \frac{k_\alpha k^\psi}{\kperp^2} \hskperp^* \left[2\rmJ_0(\besselarg) - \frac{2\rmJ_1(\besselarg)}{\besselarg}\right] \frac{w_\perp^2}{\vths^2} \frac{\dBparkperp}{B} }}, \label{eq:mflux_bpar_perp_fourier}\\
    \mfluxembpar & = -\fsa{\turbavg{ \int \rmd^3 \vec{w} \sum_{\vkperp} \Ts \frac{k_\alpha k^\psi}{\kperp^2} \hskperp^* \frac{2\rmJ_1(\besselarg)}{\besselarg} \frac{w_\perp^2}{\vths^2} \frac{\dBparkperp}{B} }}. \label{eq:mflux_bpar_em_fourier}
\end{align}

It is worth noting that the correction in \cref{eq:chi_wperp_fourer_modified} due to \cref{eq:mflux_bpar_em_fourier} is crucial for accurately capturing the size of the momentum flux arising from $\dBpar$. To see this, consider expanding the factor in the square brackets in the final term of \cref{eq:chi_wperp_fourer_modified} for $\besselarg \ll 1$. To leading order, we find that it is of size $\rmJ_0 - 2 \rmJ_1 /\besselarg = - \besselarg^2/8 + \order{\besselarg^4}$, i.e., an FLR correction to the momentum flux, as it should be. Had \cref{eq:mflux_bpar_em_fourier} not been included in \cref{eq:chi_wperp_fourer_modified}, however, one would find $\rmJ_0 - \rmJ_1 /\besselarg = 1/2 - 3 \besselarg^2/16 + \order{\besselarg^4}$, meaning that the size of this term would be overestimated by $\besselarg^{-2} \gg 1$. As such, incorrectly excluding \cref{eq:mflux_bpar_em_fourier} from the calculation of the momentum flux will give physically incorrect fluxes, and so must be included in any electromagnetic simulations regardless of the value of the plasma beta.

\subsection{Expressions for implementation in \texttt{GENE}}
\label{app:expressions_for_gene}
For the purposes of implementation in \texttt{GENE}, it will be useful to rewrite these expressions in terms of a distribution function $H_\s$, defined, in Fourier space, in terms of the non-adiabatic distribution $\hs$ as:
\begin{align}
    \hskperp = \Hskperp + \left[\rmJ_0(\besselarg) \dphipotkperp'  + \frac{2 \rmJ_1(\besselarg)}{\besselarg} \frac{w_\perp^2}{\vths^2} \frac{\Ts}{\qs} \frac{\dBparkperp}{B} \right] \frac{Z_\s e \Fs}{\Ts}.
    \label{eq:gene_df}
\end{align}
To evaluate the velocity integrals resulting from the substitution of \cref{eq:gene_df} into \cref{eq:mflux_phi_par_fourier}-\cref{eq:mflux_bpar_em_fourier}, we make use of the following identities:
\begin{align}
    \int \rmd^3 \vw \: \rmJ_0^2 \Fs &= \nss \besselgammazero, \\
    \int \rmd^3 \vw \: \frac{2 \rmJ_0 \rmJ_1}{\besselarg} \frac{w_\perp^2}{\vths^2} \Fs &= \nss \besselgammaone, \\
    \int \rmd^3 \vw \: \rmJ_0^2 \frac{\wperp^2}{\vths^2} \Fs & = \nss (\besselgammazero - \besselargint \besselgammaone), \\
    \int \rmd^3 \vw \: \frac{2\rmJ_1^2}{\besselarg^2} \frac{\wperp^4}{\vths^4} \Fs & = \nss \besselgammaone, \\
    \int \rmd^3 \vw \: \frac{2 \rmJ_0 \rmJ_1}{\besselarg} \frac{w_\perp^4}{\vths^4} \Fs &= \nss \left[\besselgammazero - (2\besselargint-1)\besselgammaone \right],
\end{align}
in which 
\begin{align}
    \besselgammazero = \rmI_{0}(\besselargint) e^{-\besselargint}, \quad \besselgammaone = \left[\rmI_{0}(\besselargint) -\rmI_{1}(\besselargint)  \right]e^{-\besselargint},
\end{align}
with $\besselargint = \kperp^2 \rhos^2/2$, $\rhos=\vths/\Omega_{\s}$, $\vths=\sqrt{2T_{\s}/m_{\s}}$, and $\rmI_0$, $\rmI_1$ are the modified Bessel functions of the first kind \cite{abramowitz72}. This allows us to obtain the following expressions for the electrostatic potential $\phipot$
\begin{align}
    \mfluxsphipar & = +\fsa{\turbavg{ \int \rmd^3 \vec{w} \sum_{\vkperp}m_{\s} c \left(\frac{I w_\parallel}{B} + \rotation R^2\right) i k_\alpha \Hskperp^* \rmJ_0(\besselarg) \dphipotkperp' }} \label{eq:mflux_phi_par_fourier_gene}\\
    & \quad + \fsa{\turbavg{\sum_{\vkperp} i k_\alpha \rotation R^2 m_{\s} c \left[ \besselgammazero \left(\frac{\qs \dphipotkperp' }{\Ts} \right)^* + \besselgammaone \left(\frac{\dBparkperp}{B}\right)^*  \right] \nss \dphipotkperp'  }}
    ,  \nonumber \\
    \mfluxsphiperp & = -\fsa{\turbavg{ \int \rmd^3 \vec{w} \sum_{\vkperp} \qs \frac{k_\alpha k^\psi}{\kperp^2} \Hskperp^* \besselarg \rmJ_1(\besselarg)\dphipotkperp' }}      \label{eq:mflux_phi_perp_fourier_gene} \\
    & \quad -\fsa{\turbavg{ \sum_{\vkperp} \frac{k_\alpha k^\psi}{\kperp^2 }  \left[\besselargint \besselgammaone \left(\frac{\qs \dphipotkperp' }{\Ts} \right)^* + 2 \besselargint \besselgammaone  \left(\frac{\dBparkperp}{B}\right)^* \right] \qs \nss \dphipotkperp'  }}; \nonumber
\end{align}
the parallel magnetic-vector potential $\dApar$
\begin{align}
    \mfluxsaparpar & = -\fsa{\turbavg{ \int \rmd^3 \vec{w} \sum_{\vkperp} m_{\s} c \left(\frac{I w_\parallel}{B} + \rotation R^2\right) i k_\alpha \Hskperp^*  \rmJ_0(\besselarg) \frac{w_\parallel \dAparkperp}{c} }} \label{eq:mflux_apar_par_fourier_gene}  \\
    & \quad -\fsa{\turbavg{\frac{I}{B} \sum_{\vkperp} i k_\alpha \left[\besselgammazero \left(\frac{\qs \dphipotkperp' }{\Ts} \right)^*  + \besselgammaone \left(\frac{\dBparkperp}{B}\right)^*  \right] \nss \Ts \dAparkperp  }}, \nonumber\\
    \mfluxsaparperp & = +\fsa{\turbavg{ \int \rmd^3 \vec{w} \sum_{\vkperp} \qs \frac{k_\alpha k^\psi}{\kperp^2} \Hskperp^*  \besselarg \rmJ_1(\besselarg) \frac{w_\parallel \dAparkperp}{c} }}, \label{eq:mflux_apar_perp_fourier_gene}\\
    \mfluxemapar & = +\fsa{\turbavg{ \int \rmd^3 \vec{w} \sum_{\vkperp} \qs \frac{k_\alpha k^\psi}{\kperp^2} \Hskperp^*  \rmJ_0(\besselarg) \frac{w_\parallel \dAparkperp}{c} }}; \label{eq:mflux_apar_em_fourier_gene}
\end{align}
and the parallel magnetic-field perturbations $\dBpar$
\begin{align}
    \mfluxsbparpar & = +\fsa{\turbavg{ \int \rmd^3 \vec{w} \sum_{\vkperp} m_{\s} c \left(\frac{I w_\parallel}{B} + \rotation R^2\right) i k_\alpha \Hskperp^*  \frac{2 \rmJ_1(\besselarg)}{\besselarg} \frac{\Ts}{\qs} \frac{w_\perp^2}{\vths^2} \frac{\dBparkperp}{B} }} \label{eq:mflux_bpar_par_fourier_gene} \\
     & \quad + \fsa{\turbavg{\sum_{\vkperp} i k_\alpha \rotation R^2 m_{\s} c \left[ \besselgammaone \left(\frac{\qs \dphipotkperp' }{\Ts} \right)^* + 2\besselgammaone \left(\frac{\dBparkperp}{B}\right)^*  \right] \frac{\nss\Ts}{\qs} \frac{\dBparkperp}{B}  }}, \nonumber\\
    \mfluxsbparperp & = +\fsa{\turbavg{ \int \rmd^3 \vec{w} \sum_{\vkperp} \Ts \frac{k_\alpha k^\psi}{\kperp^2} \Hskperp^* \left[2\rmJ_0(\besselarg) - \frac{2\rmJ_1(\besselarg)}{\besselarg}\right] \frac{w_\perp^2}{\vths^2} \frac{\dBparkperp}{B} }} \label{eq:mflux_bpar_perp_fourier_gene} \\
    & \quad +\fsa{\turbavg{ \sum_{\vkperp} \frac{k_\alpha k^\psi}{\kperp^2 } \left[ (2 \besselgammazero - \besselgammaone - 2\besselargint \besselgammaone) \left(\frac{\qs \dphipotkperp' }{\Ts} \right)^* + 2(\besselgammazero - 2 \besselargint \besselgammaone)\left(\frac{\dBparkperp}{B}\right)^* \right]  \nss \Ts \frac{\dBparkperp}{B} }}, \nonumber \\
    \mfluxembpar & = -\fsa{\turbavg{ \int \rmd^3 \vec{w} \sum_{\vkperp} \Ts \frac{k_\alpha k^\psi}{\kperp^2} \Hskperp^* \frac{2\rmJ_1(\besselarg)}{\besselarg} \frac{w_\perp^2}{\vths^2} \frac{\dBparkperp}{B} }}\label{eq:mflux_bpar_em_fourier_gene} \\
    & \quad -\fsa{\turbavg{ \sum_{\vkperp} \frac{k_\alpha k^\psi}{\kperp^2} \left[\besselgammaone \left(\frac{\qs \dphipotkperp' }{\Ts} \right)^* + 2\besselgammaone \left(\frac{\dBparkperp}{B}\right)^* \right] \nss \Ts \frac{\dBparkperp}{B} }}.
\end{align}
We note that the last two terms in \cref{eq:gene_df} made no contribution to either \cref{eq:mflux_apar_perp_fourier_gene} or \cref{eq:mflux_apar_em_fourier_gene}, since the resulting integrands are odd in the parallel velocity $\wpar$. Finally, given that throughout this paper we have been working in the \texttt{CGYRO} coordinate system, one can obtain the exact expressions that are implemented in \texttt{GENE} by the transformation $k_\alpha \rightarrow -k_{\alpha}$.

\section*{References}
\bibliography{apssamp}

\end{document}